\documentclass[reprint,superscriptaddress,amsmath,aps,showpacs,pra,longbibliography]{revtex4-2}

\pdfoutput=1
\usepackage[english]{babel}
\usepackage{bm}
\usepackage{graphicx}
\usepackage{units}
\usepackage[utf8]{inputenc}
\usepackage[colorlinks, citecolor=blue, urlcolor=blue, linkcolor=red,breaklinks]{hyperref}
\usepackage{amsmath}
\usepackage{amssymb}
\usepackage{romannum}
\usepackage{textcomp}
\DeclareUnicodeCharacter{2218}{\textdegree}
\usepackage[T1]{fontenc}
\usepackage{xspace}
\usepackage[table]{xcolor}

\begin{document}

\title{Geometrical properties of strained and twisted moir\'e heterostructures}

\author{Federico Escudero}
\email{federico.escudero@imdea.org}
\affiliation{IMDEA Nanoscience, Faraday 9, 28049 Madrid, Spain}

\author{Francisco Guinea}
\affiliation{IMDEA Nanoscience, Faraday 9, 28049 Madrid, Spain}
\affiliation{Donostia International Physics Center, Paseo Manuel de Lardiz\'{a}bal 4, 20018 San Sebasti\'{a}n, Spain}

\author{Zhen Zhan}
\email{zhenzhanh@gmail.com}
\affiliation{IMDEA Nanoscience, Faraday 9, 28049 Madrid, Spain}

\begin{abstract}

The experimental observations of many interaction-driven electronic phases in moir\'e superlattices have stimulated intense theoretical and experimental efforts to understand and engineer these correlated physics. Strain is a powerful tool for manipulating and controlling the geometrical and electronic structures of moir\'e superlattices. This review provides a comprehensive introduction to the geometry of strained moir\'e superlattices. First, starting from the linear elasticity theory, we briefly introduce the general formalism of small deformations in two-dimensional materials, and discuss the particular cases of uniaxial, shear and biaxial strain. Then, we apply the theory to twisted and strained moir\'e materials, mainly focusing on the hexagonal homobilayers, hexagonal heterobilayers and monoclinic lattices. Special moir\'e geometries, like the quasi-unidimensional patterns, square patterns and hexagonal, are theoretically predicted by manipulating the strain and twist. Finally, we review recently developed strain techniques and the special moir\'e geometries realized via these approaches. This review aims at equipping the reader with a robust understanding on the description and implementation of strain in moir\'e materials, as well as highlight some major breakthroughs in this active field.
\end{abstract}

\maketitle

\tableofcontents

\pagenumbering{arabic}
\setcounter{page}{1}
\section{Introduction}\label{sec1}

A lattice misalignment in stacked two-dimensional (2D) materials leads to moir\'e superlattices with tunable periodicity \cite{he2021moire, andrei2021marvels}. For example, bilayer graphene with a twist angle $\theta \sim 1.1^{\circ}$ generates a moir\'e pattern with a period of $\sim 14$ nm, which is termed as magic angle twisted bilayer graphene (TBG) \cite{andrei2020graphene}. Such moir\'e pattern can cause the appearance of flat bands near zero Fermi energy in the electronic band structure \cite{lopes2007graphene, bistritzer2011moire}, which are found to host an amount of topological and interaction-driven many-body states, for example, correlated insulating states \cite{cao2018correlated,xu2020correlated}, superconductivity \cite{cao2018unconventional,xia2025superconductivity,guo2025superconductivity}, ferroelectricity \cite{zheng2020unconventional}, ferromagnetism \cite{sharpe2019emergent}, integer and fractional quantum anomalous Hall states \cite{serlin2020intrinsic,xu2023observation,cai2023signatures,zeng2023thermodynamic,park2023observation}, phase transitions \cite{li2021continuous} and so on. Similar correlated phenomena have also been observed in other moir\'e heterostructures, such as twisted trilayer graphene \cite{park2021tunable, kim2022evidence}, twisted multilayer graphene \cite{zhang2022promotion}, and twisted bilayer $\mathrm{WSe_2}$ \cite{wang2020correlated, xia2025superconductivity, guo2025superconductivity}.

These rich correlated phenomena typically emerge at moir\'e scales much larger than the underlying atomic lengths, whereby the moir\'e acts as a magnifying glass of the underlying geometrical and electronic properties \cite{cosma2014moire}. This makes the moir\'e superlattices a flexible and tunable system, with plenty of tuning knobs such as twist angle, strain, external field, and substrate. Among these, strain stands out as a powerful tool to modify the geometrical and electronic properties of the stacked lattices \cite{naumis2017electronic,naumis2023mechanical,dai2019strain}. In contrast to twist-only configurations, which can only lead to the same moir\'e geometry of the underlying lattices, strain-induced deformations can lead to a plethora of different moir\'e geometries \cite{kogl2023moire, escudero2024designing}. Crucially, due to the moir\'e magnifying effect, the superlattice pattern can be significantly modified even if the layers are barely deformed \cite{cosma2014moire, escudero2024designing}.
As a result, strains can strongly change the moir\'e size, the superlattice symmetry, and by extension, the electronic properties of the system. 

While strain in moir\'e superlattices is well known to arise naturally during their fabrication \cite{kazmierczak2021strain}, recent experimental advancements have opened the possibility of actively induce and control external strains in the system \cite{kapfer2023programming, pena2023moire, ou2025continuous}. A precise control on both twist and strain greatly magnifies the tunability of moir\'e superlattices, paving the way for potentially novel correlated phenomena. Since the electronic properties of moir\'e systems depend largely on the electronic modulation due to the moir\'e potential, it becomes imperative to understand precisely how the geometrical properties of moir\'e heterostructures depend on the interplay between twist and strain. Among different types of strain, uniaxial heterostrain is the most common type, and has been observed in many experimental samples \cite{huder2018electronic,kerelsky2019maximized,xie2019spectroscopic,mesple2021heterostrain,wang2023unusual}. Recently, under some new well-developed techniques, shear heterostrain is achieved, resulting in deformed moir\'e patterns \cite{yu2024twist,carrasco2025}.

Although recent works have already provide a review of the strain effect and implementation in moir\'e heterostructures \cite{he2021moire, hou2025strain, zhai2025twistronics, trambly2026electronic}, there is yet no detailed review of the geometrical description. Namely, on how does the geometry of the moir\'e patterns depend on the interplay between twist and strain, and in particular, on the different types of strain configurations that can be engineered in experimental setups. The goal of this review is to cover this gap by offering an overview of the recent advances on the treatment and implementation of strain in moir\'e heterostructures. To make this work comprehensive, we start by first revisiting the linear elasticity theory, adapted to common in-plane stress and strain in 2D materials. Next, we apply the linear elasticity theory to the twisted and strained moir\'e systems, and describe how many different moir\'e geometries can be archived by tuning both the twist and strain. Finally, we discuss recent major breakthroughs in implementing strain in moir\'e heterostructures, and show experimental examples of particular moir\'e geometries. 

\section{Linear elasticity in two-dimensional materials}

In moir\'e heterostructures made up of 2D materials (one-atom thickness), the most relevant stress-induced deformations are planar, i.e., in-plane displacements that act on the 2D plane of the material \cite{pereira2009tight}. However, there can also be out-of-plane displacements in the atomic positions (e.g., corrugations). For instance, thermal ripples are well known to appear naturally in graphene at room temperatures \cite{meyer2007structure, fasolino2007intrinsic}. Moreover, out-of-plane corrugations emerge naturally when one stacks two or more 2D materials with a lattice mismatch (induced, e.g., by twist or strain), as the system can reduce its energy by increasing the distance at the energetically costly stacking regimes \cite{uchida2014atomic, nam2017lattice, koshino2018maximally, cazeaux2023relaxation}. 

In this review, we will simplify the problem by considering only in-plane stresses. This is a safe approximation as our focus is only to describe the geometrical properties of the emergent in-plane moir\'e patterns. We label the 2D materials plane with the coordinates $x$ and $y$. For self-consistence, and to set up the ground for the following sections, in this section we briefly review the theory of linear elasticity \cite{landau2012theory, gurtin1982introduction, gurtin2010mechanics}, adapted to in-plane stress and strain in 2D materials.

\subsection{Displacement field, strain and stress}

In elasticity theory, the relevant quantity that identifies a deformation of a material is the displacement field $\mathbf{u}$, defined as \cite{landau2012theory, gurtin2010mechanics}
\begin{equation}
\mathbf{u}\left(\mathbf{r}\right)=\mathbf{r}'-\mathbf{r},
\end{equation}
where $\mathbf{r}'=\left(x',y'\right)$ and $\mathbf{r}=\left(x,y\right)$ are the deformed and undeformed position of the material, respectively. In general, the displacement vector could also depend on the time $t$, i.e., $\mathbf{u}\left(\mathbf{r},t\right)$. Here, we simplify the problem and consider time-independent deformations, as is usually the case in two-dimensional materials \cite{vozmediano2010gauge, guinea2012strain, amorim2016novel, si2016strain}. Using that $d\mathbf{r}'=d\mathbf{r}+d\mathbf{u}$ and $d\mathbf{u}=\nabla\mathbf{u}d\mathbf{r}$ (where $\nabla$ is the gradient operator), the infinitesimal length $d\ell'=\left|d\mathbf{r}'\right|$ of the deformed material reads \cite{landau2012theory}
\begin{equation}
d\ell'{}^{2}=d\ell^{2}+2\mathcal{E}d\mathbf{r}\cdot d\mathbf{r},
\end{equation}
where $d\ell=\left|d\mathbf{r}\right|$ is the infinitesimal length of the undeformed material and $\mathcal{E}$ is the strain tensor defined as
\begin{equation}
\mathcal{E}=\frac{\nabla\mathbf{u}+\left(\nabla\mathbf{u}\right)^{\mathrm{T}}+\nabla\mathbf{u}\left(\nabla\mathbf{u}\right)^{\mathrm{T}}}{2}.
\end{equation}
In index notation
\begin{equation}
\epsilon_{ij}=\frac{1}{2}\left(\frac{\partial u_{i}}{\partial x_{j}}+\frac{\partial u_{j}}{\partial x_{i}}+\frac{\partial u_{i}}{\partial x_{j}}\frac{\partial u_{j}}{\partial x_{i}}\right),
\end{equation}
where $\epsilon_{ij}\equiv\left(\mathcal{E}\right)_{ij}$ are the $i,j=x,y$ (in-plane) components. The strain tensor is a measure of the change in infinitesimal lengths due to deformations. 

In most practical cases, the deformation of a material is relatively small \cite{amorim2016novel, kazmierczak2021strain, mesple2021heterostrain}. Even in very strong two-dimensional materials like graphene --which can stand up to about 20\% strain \cite{lee2008measurement, papageorgiou2017mechanical}-- most experimental deformations are at most within the range of $1\%$ and $2\%$ \cite{frisenda2017biaxial, qiao2018twisted, mesple2021heterostrain, kazmierczak2021strain}. We refer to these as \emph{small deformations}. In such case one can neglect the terms $\mathcal{O}\left(\nabla\mathbf{u}\right)^{2}$ and take \cite{landau2012theory}
\begin{equation}
\mathcal{E}\approx\frac{\nabla\mathbf{u}+\left(\nabla\mathbf{u}\right)^{\mathrm{T}}}{2}\Rightarrow\epsilon_{ij}=\frac{1}{2}\left(\frac{\partial u_{i}}{\partial x_{j}}+\frac{\partial u_{j}}{\partial x_{i}}\right).
\label{eq:strain_linear}
\end{equation}
This is, indeed, what is usually taken as the strain tensor. 

The linear theory of elasticity follows the regime of small deformations by keeping all the relevant quantities only up to leading order (linear) in the strain tensor \cite{gurtin2010mechanics}. An important quantity, besides the strain tensor, is the so-called stress tensor $\mathcal{T}$. It is defined so that the tension $\mathbf{t}$ (force per unit length in two-dimensions), at a surface point with normal vector $\mathbf{n}$, is given by the Cauchy's theorem $\mathbf{t}=\mathcal{T}\mathbf{n}$ \cite{gurtin2010mechanics} (for in-plane stress and strain, the normal vector $\mathbf{n}$ is either of the two in-plane directions, i.e., $\mathbf{n}=\mathbf{e}_{x}$ or $\mathbf{n}=\mathbf{e}_{y}$, or any combination of them). Within the linear theory of elasticity, the stress tensor $\mathcal{T}$ is related to the strain tensor through a generalized Hook's law \cite{gurtin2010mechanics}
\begin{equation}
\mathcal{T}=\mathbb{C}\mathcal{E},
\label{eq:Cauchy}
\end{equation}
where $\mathbb{C}$ is the fourth-order elasticity tensor (also called stiffness or compliance tensor \cite{pereira2009tight}). Equation \eqref{eq:Cauchy} gives the stress-strain relation that connects forces to deformations. In practice, the elasticity tensor $\mathbb{C}$ is determined by constitutive equations. 

For isotropic materials the relation $\mathcal{T}=\mathbb{C}\mathcal{E}$ becomes a linear isotropic function of the strain tensor; its most general representation reads \cite{gurtin1982introduction}
\begin{equation}
\mathcal{T}=2\mu\mathcal{E}+\lambda\left(\mathrm{tr}\mathcal{E}\right)\mathbb{I},\label{eq:Stress}
\end{equation}
where $\mu$ and $\lambda$ are known as the Lam\'e coefficients, and $\mathbb{I}$ is the $2\times2$ identity matrix. One can then invert Eq. \eqref{eq:Stress} and write the strain tensor in terms of the stress tensor as \cite{gurtin2010mechanics}
\begin{equation}
\mathcal{E}=\frac{1}{2\mu}\left[\mathcal{T}-\frac{\lambda}{2\left(\mu+\lambda\right)}\mathrm{tr}\left(\mathcal{T}\right)\mathbb{I}\right].
\label{eq:Strain}
\end{equation}
Equations \eqref{eq:Stress} and \eqref{eq:Strain} constitute the fundamental mechanical relations of linear elasticity theory in 2D materials. 

\subsection{Uniaxial, shear and biaxial strain}\label{sec:Strain_types}

\begin{figure}[t]
\centering
    \includegraphics[width=0.9\linewidth]{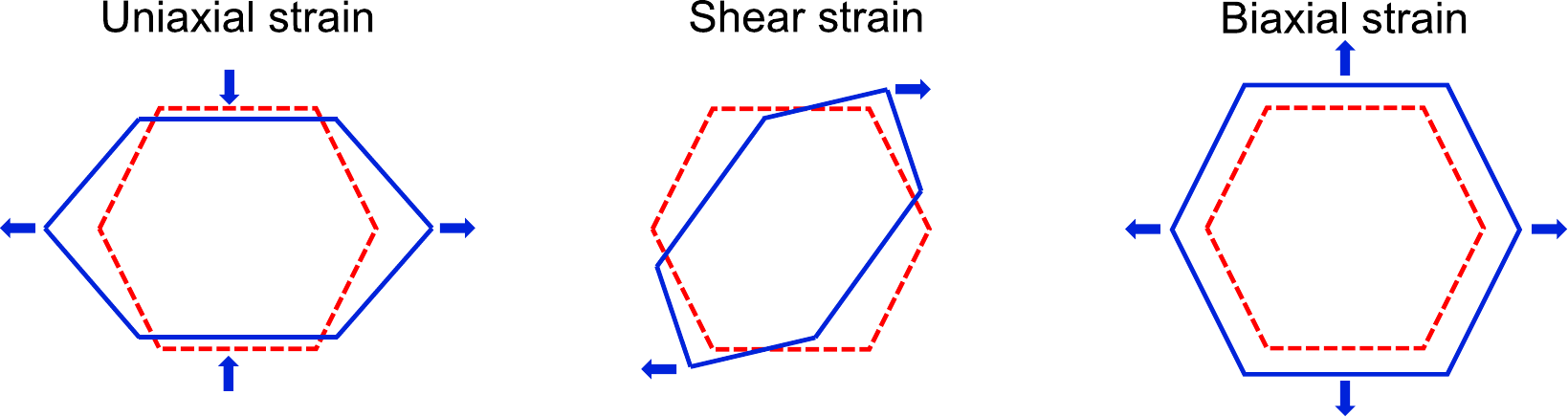}
    \caption{Schematic effect of three common types of strain in moir\'e systems: uniaxial, shear and biaxial. The red dashed-line shows the undeformed hexagon and the blue solid-line shows the deformation of each strain type (uniaxial and shear, both with direction along the x-axis).}\label{fig:Strain_types}
\end{figure}

In what follows we will limit our discussion to homogeneous (position-independent) strain (for reviews of non-homogeneous strain, see e.g. Refs. \cite{amorim2016novel, naumis2017electronic}). The three experimentally-relevant types of stress in 2D materials are uniaxial, shear, and biaxial, shown in Figure \ref{fig:Strain_types}.

\subsubsection{Uniaxial strain}
A uniaxial force, either tension or compression, acts along a fixed axis. Taking such axis as the $x$-axis, the surface-tension is given by $\mathbf{t}=\sigma_u\mathbf{e}_{x}$, where $\sigma_u$ is the force magnitude per unit length. By the Cauchy's relation, the stress tensor reads 
\begin{equation}
\mathcal{T}=\left(\begin{array}{cc}
\sigma_u & 0\\
0 & 0
\end{array}\right).
\end{equation}
The uniaxial strain then becomes [cf. Eq. \eqref{eq:Strain}]
\begin{equation}
\mathcal{E}=\left(\begin{array}{cc}
\epsilon_u & 0\\
0 & -\nu\epsilon_u
\end{array}\right),
\label{eq:uniaxial_strain}
\end{equation}
where
\begin{equation}
\epsilon_u=\frac{\sigma_u}{4\mu}\frac{2\mu+\lambda}{\mu+\lambda},\quad\nu=\frac{\lambda}{2\mu+\lambda},
\end{equation}
are the longitudinal strain magnitude and Poisson's ratio, respectively \cite{landau2012theory, gurtin1982introduction}. A related quantity is the Young's modulus $E$, given by the ratio between the longitudinal strain and the stress
\begin{equation}
E=\frac{\sigma_u}{\epsilon_u}=4\mu\frac{\mu+\lambda}{2\mu+\lambda}
\end{equation}
Since $2\mathcal{E}=\nabla\mathbf{u}+\left(\nabla\mathbf{u}\right)^{\mathrm{T}}$ (Eq. \eqref{eq:strain_linear}), the displacement field for the uniaxial strain reads 
\begin{equation}
\mathbf{u}(\mathbf{r})=\epsilon_u (x\mathbf{e}_{x}-\nu y\mathbf{e}_{y} ),
\end{equation}
so the Poisson's ratio measures the contraction lateral to the direction of the applied stress \cite{gurtin2010mechanics}. The magnitude of such contraction is determined by the material-dependent value of the Poisson's ratio (e.g., $\nu\sim 0.16$ in graphene \cite{blakslee1970elastic, pereira2009tight}).

If the uniaxial stress acts along a direction $\phi$ with respect to the $x$-axis, then the uniaxial strain becomes \cite{pereira2009tight, bi2019designing}
\begin{align}
\mathcal{E} & =R\left(\phi\right)\left(\begin{array}{cc}
\epsilon_u & 0\\
0 & -\nu\epsilon_u
\end{array}\right)R\left(-\phi\right)\nonumber \\
 & =\epsilon_u\left[\begin{array}{cc}
\cos^{2}\phi-\nu\sin^{2}\phi & \left(1+\nu\right)\sin\phi\cos\phi\\
\left(1+\nu\right)\sin\phi\cos\phi & \sin^{2}\phi-\nu\cos^{2}\phi
\end{array}\right],
\label{eq:strain_uniaxial_phi}
\end{align}
where $R\left(\phi\right)$ is the rotation matrix
\begin{equation}
R\left(\phi\right)=\left(\begin{array}{cc}
\cos\phi & -\sin\phi\\
\sin\phi & \cos\phi
\end{array}\right).
\end{equation}
Equation \eqref{eq:strain_uniaxial_phi} follows by first rotating the system by $-\phi$ to bring the stress along the $x$-axis, then strain it according to Eq. \eqref{eq:uniaxial_strain}, and finally rotate it back by $\phi$.

\subsubsection{Shear strain}
Shear forces act perpendicularly to the normal of the surface. A shear tension perpendicular to the $x$-axis is given by $\mathbf{t}=\sigma_s\mathbf{e}_{y}$, and implies a symmetric shear stress 
\begin{equation}
\mathcal{T}=\left(\begin{array}{cc}
0 & \sigma_s\\
\sigma_s & 0
\end{array}\right).
\end{equation}
Since $\mathrm{tr}\left(\mathcal{T}\right)=0$, the shear strain is traceless and proportional to $\mathcal{T}$
\begin{equation}
\mathcal{E}=\left(\begin{array}{cc}
0 & \epsilon_s\\
\epsilon_s & 0
\end{array}\right),
\label{eq:strain_shear}
\end{equation}
where $\epsilon_s=\sigma_s/2\mu$ is the shear strain magnitude. The Lam\'e coefficient $\mu$ is thus also called the \emph{shear modulus} \cite{landau2012theory, gurtin2010mechanics}. The shear displacement vector reads $\mathbf{u}=2\epsilon_s x\mathbf{e}_{y}$. If the shear acts, in general, perpendicular to a direction $\phi$ (relative to the $x$-axis), the shear strain reads \cite{kogl2023moire, escudero2024designing}
\begin{align}
\mathcal{E} & =R\left(\phi\right)\left(\begin{array}{cc}
0 & \epsilon_s\\
\epsilon_s & 0
\end{array}\right)R\left(-\phi\right)\nonumber \\
 & =\epsilon_s\left(\begin{array}{cc}
-\sin2\phi & \cos2\phi\\
\cos2\phi & \sin2\phi
\end{array}\right).
\label{eq:strain_shear_phi}
\end{align}
Note that regardless of the direction, the shear strain is always traceless. This means that shear forces deform the materials without changing their area. 

\subsubsection{Biaxial strain}
Biaxial forces are those that contract or expand a body uniformly, without changing its shape. It corresponds to a tension $\mathbf{t}=\sigma_b\left(\mathbf{e}_{x}+\mathbf{e}_{y}\right)$, which implies a spherical biaxial stress tensor $\mathcal{T}=\sigma_b\mathbb{I}$. The biaxial strain is then also spherical
\begin{equation}
\mathcal{E}=\left(\begin{array}{cc}
\epsilon_b & 0\\
0 & \epsilon_b
\end{array}\right),
\label{eq:strain_biaxial}
\end{equation}
where
\begin{equation}
\epsilon_b=\frac{1}{2}\frac{\sigma_b}{\mu+\lambda}
\end{equation}
is the biaxial strain magnitude. The ratio between the stress and the strain (Young's modulus) is given by $\sigma_b/\epsilon_b=2\left(\mu+\lambda\right)$. Note that the biaxial strain is independent of the direction because it contracts (or expands) the material uniformly in each direction.

\subsection{Deformation of bonds}

In the absence of external body forces, a displacement field $\mathbf{u}$ with constant gradient $\nabla\mathbf{u}$ is always a static solution to the equation of motion \cite{landau2012theory, gurtin1982introduction, gurtin2010mechanics}. A constant gradient $\nabla\mathbf{u}$ corresponds to a homogeneous (position-independent) strain tensor [cf. Eq. \eqref{eq:strain_linear}], as considered in Section \ref{sec:Strain_types}. The displacement field $\mathbf{u}$ can then be written in terms of the strain tensor as
\begin{equation}
\mathbf{u}=\mathcal{E}\mathbf{r}.
\end{equation}
In index notation, $u_{i}=\epsilon_{ij}x_{j}$, where repeated indices are to be summed over. Indeed, taking the gradient of $\mathbf{u}=\mathcal{E}\mathbf{r}$, and using the symmetry of the strain tensor, directly leads to Eq. \eqref{eq:strain_linear}. Here we have fixed, without loss of generality, the boundary condition by taking the displacement equal to zero at the origin (i.e., $\mathbf{u}=\mathbf{0}$ when $\mathbf{r}=0$). At small deformations, the deformed bonds (or lengths) of a material then become \cite{pereira2009tight}
\begin{equation}
\mathbf{r}'=\left(\mathbb{I}+\mathcal{E}\right)\mathbf{r}.
\label{eq:deformation}
\end{equation}
This is the fundamental relation that determines the (small) deformation of any bond at undeformed position $\mathbf{r}$. In particular, for a 2D crystal with lattice vectors $\mathbf{a}_{1}$, $\mathbf{a}_{2}$, and basis vector $\boldsymbol{\delta}$, all atomic positions $\mathbf{r}=n_{1}\mathbf{a}_{1}+n_{2}\mathbf{a}_{2}+\boldsymbol{\delta}$ (with $n_{1}$ and $n_{2}$ integers) are transformed by the strain to $\mathbf{r}\rightarrow\left(\mathbb{I}+\mathcal{E}\right)\mathbf{r}$.

\section{Twisted and strained moir\'e geometries}\label{sec:twist_strain}

Two-dimensional materials stacked in a multilayer configuration, subject to relative twist and strain, create a lattice mismatch between the layers and, consequently, a moir\'e pattern \cite{moon2013optical, bi2019designing, kogl2023moire, escudero2024designing}. Such structures are thus referred as \emph{moir\'e heterostructures} \cite{ andrei2021marvels, kennes2021moire}. In this review we will mostly focus on the simplest case of two layers stacked in a bilayer configuration \cite{andrei2020graphene}.  To simplify the discussion, we start by considering the simplest (but most common) case of two equal hexagonal lattices. Generalizations to different layers are discussed in subsequent sections.

\subsection{Hexagonal homobilayers}\label{sec:homobilayers}

We describe the honeycomb lattices by a set of lattice vectors \cite{castro2009electronic}
\begin{equation}
\mathbf{a}_{1}=a\left(1,0\right),\quad\mathbf{a}_{2}=a\left(1/2,\sqrt{3}/2\right),
\end{equation}
and a basis of two atoms at positions $\mathbf{d}_{1}=\left(0,0\right)$ and $\mathbf{d}_{2}=\left(0,d\right)$. Here $a$ is the lattice constant (e.g., $a\approx2.46\:\textrm{Å}$ in monolayer graphene) and $d=a/\sqrt{3}$ is the
interatomic spacing. The atomic basis defines the two sublattices $A$ and $B$ of the honeycomb structure. The three nearest neighbors of an atom in the sublattice $A$ are 
\begin{equation}
\boldsymbol{\delta}_{1}=d\left(0,1\right),\quad\boldsymbol{\delta}_{2}=\boldsymbol{\delta}_{1}-\mathbf{a}_{2},\quad\boldsymbol{\delta}_{3}=\boldsymbol{\delta}_{1}+\mathbf{a}_{1}-\mathbf{a}_{2}.
\end{equation}
The reciprocal lattice vectors $\mathbf{b}_{i}$, which satisfy $\mathbf{a}_{i}\cdot\mathbf{b}_{j}=2\pi\delta_{ij}$, read 
\begin{equation}
\mathbf{b}_{1}=b\left(\sqrt{3}/2,-1/2\right),\quad\mathbf{b}_{2}=b\left(0,1\right),
\label{eq:reciprocal}
\end{equation}
where $b=4\pi/\sqrt{3}a$.

To describe the geometry of the twisted and strained honeycomb layers, is convenient to consider the effect of twist and strain separately. Specifically, we consider the effect of first twisting and then straining \cite{bi2019designing, sinner2023strain, kogl2023moire, escudero2024designing}. A twist by an angle $\theta$ simply rotates the atomic bonds as $\mathbf{r}\rightarrow R\left(\theta\right)\mathbf{r}$. A further strain on the rotated system will then, according to Eq. \eqref{eq:deformation}, transform the atomic positions as $\mathbf{r}'=\left(\mathbb{I}+\mathcal{E}\right)R\left(\theta\right)\mathbf{r}$. 

The moir\'e pattern emerges when the two layers are twisted and strained differently. This configuration with strain is known as \textit{heterostrain}. The hexagonal lattice vectors $\mathbf{a}_{i}$ ($i=1,2$) in each layer transform as
\begin{equation}
\mathbf{a}_{i,\pm}=\left(\mathbb{I}+\mathcal{E}_{\pm}\right)R\left(\theta_{\pm}\right)\mathbf{a}_{i},
\label{eq:ai_strain}
\end{equation}
where $+$ and $-$ refer to the top and bottom layers, respectively. Within the small deformation limit one often keeps only the linear terms in the twist and strain, so that \cite{bi2019designing}
\begin{align}
\left(\mathbb{I}+\mathcal{E}\right)R\left(\theta\right)  \approx\mathbb{I}+\left(\begin{array}{cc}
\epsilon_{xx} & \epsilon_{xy}-\theta\\
\epsilon_{xy}+\theta & \epsilon_{yy}
\end{array}\right)
 =\mathbb{I}+\mathcal{R}\left(\theta\right)+\mathcal{E},
\label{eq:strain_general}
\end{align}
where the symmetric part $\mathcal{E}$ gives the strain effect and the antisymmetric part $\mathcal{R}\left(\theta\right)$ gives the twist effect. In general, the three components of the strain tensor can be decomposed as a mixture of biaxial and shear strains \cite{halbertal2021moire, kogl2023moire}
\begin{equation}
\mathcal{E}=\epsilon_{b}\mathbb{I}+\epsilon_{s}S\left( \phi \right),
\end{equation}
where 
\begin{equation}
\epsilon_{b}=\frac{\epsilon_{xx}+\epsilon_{yy}}{2},\qquad\epsilon_{s}=\sqrt{\left(\frac{\epsilon_{xx}-\epsilon_{yy}}{2}\right)^{2}+\epsilon_{xy}^{2}},
\label{eq:strain_equiv}
\end{equation}
are the biaxial and shear strain magnitudes, respectively. The shear matrix $S\left( \phi \right)$ is defined as
\begin{equation}
S\left( \phi \right)=\left(\begin{array}{cc}
\sin\phi & \cos\phi\\
\cos\phi & -\sin\phi
\end{array}\right),
\label{eq:shear_matrix}
\end{equation}
where $\cos\phi=\epsilon_{xy}/\epsilon_{s}$. In this parameterization the uniaxial strain becomes a mixture of biaxial and shear strain \cite{kogl2023moire} (more details in Section \ref{sec:Shear_and_Biaxial}).

A symmetric bilayer configuration is to twist and strain the two layers with the same magnitude but in opposite direction \cite{bi2019designing, sinner2023strain, escudero2024designing}, so that $\theta$$_{\pm}=\pm\theta/2$ and $\mathcal{E}_{\pm}=\pm\mathcal{E}/2$. But there are, of course, many other possibilities. For instance, one layer could be only rotated while the other layer is only strained \cite{kogl2023moire}. For conciseness and simplicity, in this section we will consider the symmetric configuration. The presented analysis is, however, general and does not depend on the particular twist and strain in each layer.

In general, for arbitrary twist and strain, the moir\'e pattern is not commensurate \cite{bi2019designing, escudero2024designing}. Commensuration only occurs at particular twist and strain for which one can find a superlattice vector common to both layers \cite{shallcross2010electronic, moon2013optical, shi2020large, wang2025}. Here we shall not pursue the problem of finding commensurate solutions, and consider arbitrary twist and strain regardless of whether they correspond to a commensurate or incommensurate moiré. As we will see, the geometrical properties of the moir\'e pattern can be well defined and studied even if the system is technically incommensurate \cite{koshino2015interlayer}. In fact, in the relevant regime of moir\'e scales much larger than the atomic lengths, the electronic properties become insensitive to the commensurability of the system \cite{bistritzer2011moire}. Effective continuum models, widely employed to study the electronic properties of large moir\'e heterostructures, can be well defined even if the system is incommensurate \cite{koshino2015interlayer, bi2019designing}. The commensuration is, however, fundamental when one needs to compute the band structure by more realistic methods, such as atomistic tight-binding \cite{suarez2010flat, long2022atomistic,wang2025} or density functional theory (DFT) simulations \cite{cantele2020structural, leconte2022relaxation}.

\begin{figure}[t]
\centering
    \includegraphics[width=0.9\linewidth]{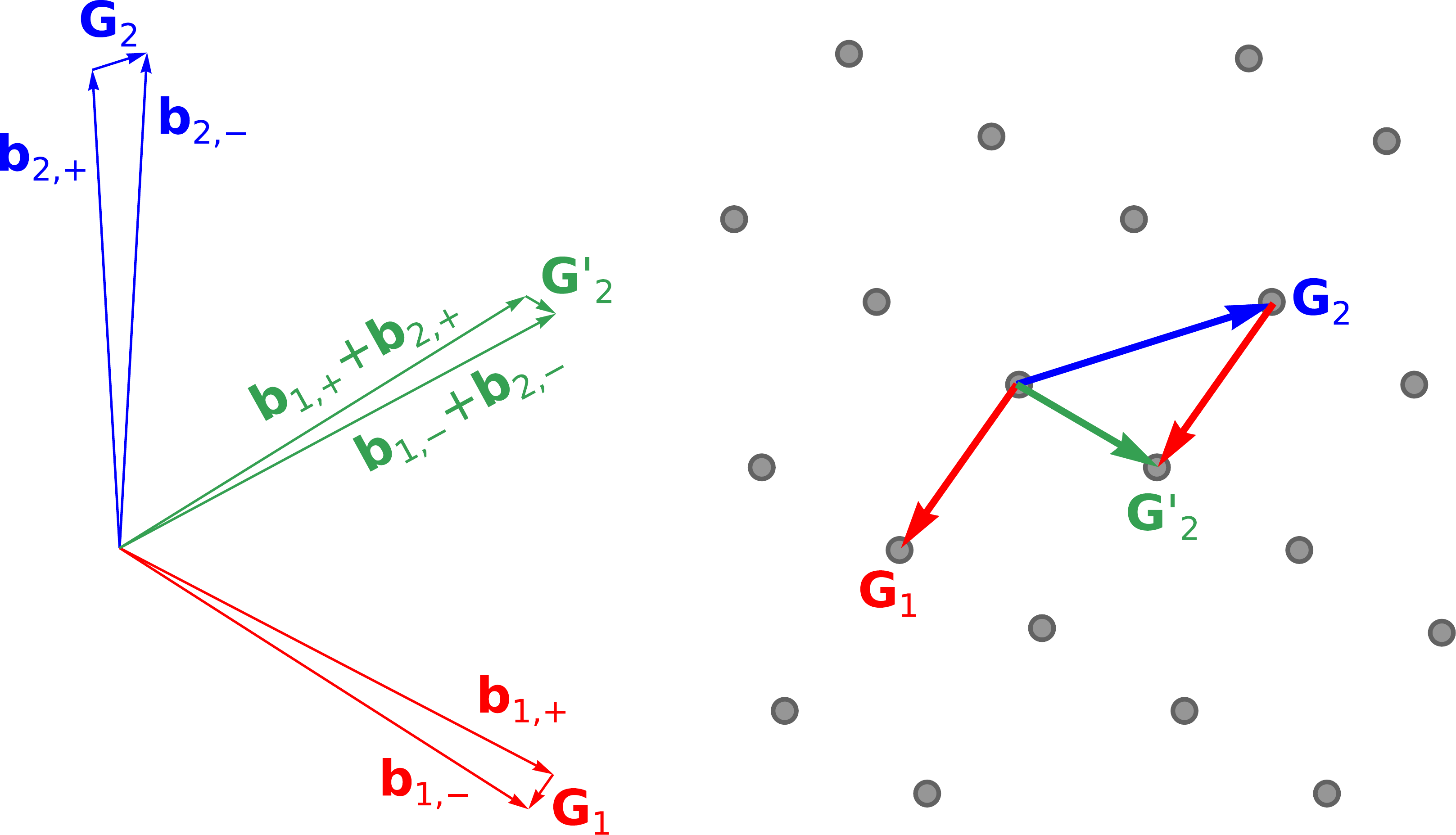}
    \caption{Strain-dependent construction of moir\'e vectors. The left panels shows the deformed reciprocal vectors $\mathbf{b}_{i,\pm}$ in two honeycomb lattices subject to a twist $\theta=5^{\circ}$ and uniaxial strain with magnitude $\epsilon=5\%$ and direction $\phi=60^{\circ}$. Taking the difference between the deformed reciprocal vectors does not yield, in this case, the smallest moir\'e vectors $\mathbf{G}_{i}$. A smaller moir\'e vector $\mathbf{G}'_{2}=\mathbf{G}_{1}+\mathbf{G}_{2}$ is obtained by taking the difference between the deformed reciprocal vectors $\mathbf{b}'_{2,\pm}=\mathbf{b}_{1,\pm}+\mathbf{b}_{2,\pm}$. The right panel shows the superlattice spanned by the moir\'e vectors. Adapted under the terms of the CC BY license from Ref. \cite{escudero2024designing}. Copyright (2024) by the American Physical Society.}\label{fig:Moire_vectors}
\end{figure}

The geometrical properties of the moir\'e pattern are more easily described by working in reciprocal space \cite{bi2019designing, escudero2024designing}. The twisted and strained reciprocal lattice vectors $\mathbf{b}_{i,\pm}$ in each layer follow from Eq. \eqref{eq:ai_strain} and the condition $\mathbf{a}_{i,\pm}\cdot\mathbf{b}_{j,\pm}=2\pi\delta_{ij}$, which implies
\begin{align}
\mathbf{b}_{i,\pm}  =\left(\mathbb{I}+\mathcal{E}_{\pm}\right)^{-1}R\left(\theta_{\pm}\right)\mathbf{b}_{i}
 \approx\left(\mathbb{I}-\mathcal{E}_{\pm}\right)R\left(\theta_{\pm}\right)\mathbf{b}_{i}.
\end{align}
The last approximation holds within the small deformation regime \cite{bi2019designing}. The \emph{moir\'e vectors} of the moir\'e pattern can be formally defined as the set of vectors $\mathbf{G}_{M}$ that satisfy
\begin{equation}
e^{i\mathbf{G}_{M}\cdot\mathbf{R}_{M}}=1,
\end{equation}
where $\mathbf{R}_{M}$ are the minimal union of the lattice vectors $\mathbf{R}_{\pm}=n_{1}\mathbf{a}_{1,\pm}+n_{2}\mathbf{a}_{2,\pm}$ in each layer (with $n_{1}$ and $n_{2}$ integers). In a commensurate structure $\mathbf{R}_{M}$ reduce to the superlattice vectors at which $\mathbf{R}_{M}=\mathbf{R}_{\pm}$ \cite{shallcross2010electronic}, but in an incommensurate structure $\mathbf{R}_{M}$ is generally given by the lattice points at which the difference between $\mathbf{R}_{+}$ and $\mathbf{R}_{-}$ is minimum, which always occurs around the AA stacking regimes \cite{moon2013optical}. The \textit{primitive moir\'e vectors} $\mathbf{G}_{i}$ ($i=1,2$) are the two smallest vectors $\mathbf{G}_{M}$ from which all others can be obtained by translation. In the case of only a twist, the primitive moir\'e vectors are given by the difference between the reciprocal lattice vectors \cite{moon2013optical, koshino2015interlayer, koshino2018maximally}
\begin{equation}
\mathbf{G}_{i}=\mathbf{b}_{i,-}-\mathbf{b}_{i,+}.\label{eq:moire_vectors}
\end{equation}
However, this is not necessarily the case in the presence of strain. Indeed, the deformation of each lattice may be such that other combination between $\mathbf{b}_{i,+}$ and $\mathbf{b}_{i,-}$ may yield a smaller moir\'e vector \cite{escudero2024designing} (see Figure \ref{fig:Moire_vectors}). Accounting for these different constructions of the moir\'e vectors is essential to elucidate the wide range of moir\'e patterns that can be formed by strain. The analysis can be simplified by noting that the different geometrical constructions of the primitive moir\'e vectors must be related by the hexagonal symmetry of the underlying lattices. This means that it is sufficient to study, as usually done, only the geometrical properties of the moir\'e vectors given by Eq. \eqref{eq:moire_vectors}, keeping in mind that all other constructions are related by $60^{\circ}$ translations of the strain parameters \cite{escudero2024designing}. 

For stacking configurations made of two homobilayers, the moir\'e vectors given by Eq. \eqref{eq:moire_vectors} are determined by a unique transformation $\mathbf{T}$ acting on the reciprocal vectors \cite{sinner2023strain}, $\mathbf{G}_{i}=\mathbf{T}\mathbf{b}_{i}$, where (for the symmetrical configuration)
\begin{equation}
\mathbf{T}=\left(\mathbb{I}+\mathcal{E}/2\right)\mathrm{R}\left(-\theta/2\right)-\left(\mathbb{I}-\mathcal{E}/2\right)\mathrm{R}\left(\theta/2\right).
\label{eq:T}
\end{equation}
From the moir\'e vectors one can describe the geometrical properties of the moir\'e pattern by two parameters: their relative angle $\beta$ and their relative length \cite{kogl2023moire, escudero2024designing}. For moir\'e patterns arising from only a relative rotation of the honeycomb lattices (i.e., with no strain), $\beta=120^{\circ}$ is always the same as the angle between the reciprocal vectors $\mathbf{b}_{i}$. This changes in the presence of strain due to the lattice deformation. The angle $\beta$ can be obtained through the relation $\mathbf{G}_{1}\cdot\mathbf{G}_{2}=\left|\mathbf{G}_{1}\right|\left|\mathbf{G}_{2}\right|\cos\beta$. Using that $\mathbf{G}_{i}=\mathbf{T}\mathbf{b}_{i}$ and the property \cite{gurtin2010mechanics} $\mathbf{T}\mathbf{b}_{1}\cdot\mathbf{T}\mathbf{b}_{2}=\left(\mathbf{T}^{\mathrm{T}}\mathbf{T}\right)\mathbf{b}_{1}\cdot\mathbf{b}_{2}$, where $\mathbf{T}^{\mathrm{T}}$ is the transpose of $\mathbf{T}$, it follows that $\beta$ is uniquely determined by the symmetric transformation
$\mathbf{F}=\mathbf{T}^{\mathrm{T}}\mathbf{T}$ acting on the reciprocal vectors \cite{escudero2024designing} 
\begin{equation}
\cos\beta=\frac{\mathbf{F}\mathbf{b}_{1}\cdot\mathbf{b}_{2}}{\sqrt{\left(\mathbf{F}\mathbf{b}_{1}\cdot\mathbf{b}_{1}\right)\left(\mathbf{F}\mathbf{b}_{2}\cdot\mathbf{b}_{2}\right)}}.
\label{eq:cosB}
\end{equation}
Here, $\sqrt{\mathbf{F}\mathbf{b}_{i}\cdot\mathbf{b}_{i}}=\left|\mathbf{G}_{i}\right|$ gives the length of the moir\'e vectors, while $\mathbf{F}\mathbf{b}_{1}\cdot\mathbf{b}_{2}=\mathbf{G}_{1}\cdot\mathbf{G}_{2}$ gives their projection. It is instructive to separate $\mathbf{F}=\mathbf{F}_{0}+\mathbf{F}_{\epsilon}$, where $\mathbf{F}_{0}$ is the contribution due to pure rotations, and $\mathbf{F}_{\epsilon}$ is the contribution due to the combination of rotation and strain
\begin{align}
\mathbf{F}_{0} & =4\sin^{2}\frac{\theta}{2}\mathbb{I},\label{eq:F_0}\\
\mathbf{F}_{\epsilon} & =\sin\theta\left(\begin{array}{cc}
-2\epsilon_{xy} & \epsilon_{xx}-\epsilon_{yy}\\
\epsilon_{xx}-\epsilon_{yy} & 2\epsilon_{xy}
\end{array}\right)+\mathcal{E}^{2}\cos^{2}\frac{\theta}{2}.\label{eq:Fe}
\end{align}
The transformation by $\mathbf{F}_{0}$ is a spherical tensor (i.e., a tensor proportional to the identity matrix $\mathbb{I}$), so rotations alone do not change the angle of the moir\'e vectors. However, in the presence of strain the non-zero transformation $\mathbf{F}_{\epsilon}$ can modify the angle $\beta$ in Eq. \eqref{eq:cosB}, and thus the geometry of the moir\'e patterns. Note that the expression of $\mathbf{F}_{\epsilon}$ depends the interplay between twist and strain; in particular, $\mathbf{F}_{\epsilon}$ is not a spherical tensor only for strain configurations that deform the lattices unit cell (in other words, for non-spherical strain tensors $\mathcal{E}$). The last term in $\mathbf{F}_{\epsilon}$ describes the possibility of obtaining moir\'e patterns purely by strain \cite{san2012non}.

Significant modifications of the moir\'e vectors by the strain can be archived even at very small strain magnitudes, which may only barely distort the underlying lattices \cite{kogl2023moire, escudero2024designing}. This is because the observed moir\'e generally acts as a magnifying glass that enhances any small lattice mismatch between the layers \cite{cosma2014moire}. One can see this in a simple example of both twist and uniaxial heterostrain with magnitude $\epsilon$ along the $x$ direction. In such case the angle between the moir\'e vectors, to leading order in $\theta$ and $\epsilon_u$, is given by \cite{escudero2024designing}
\begin{equation}
\cos\beta\approx-\frac{1}{2}+\frac{3\sqrt{3}}{8}\left(\nu+1\right)\frac{\epsilon_u}{\theta}.
\end{equation}
The dependence $\sim\epsilon_u/\theta$ reflects that the smaller the twist angle, the smaller the strain magnitude needed to modify the moir\'e geometry. As a result, for low twist angles and experimentally relevant values $\epsilon\lesssim10\%$, one can in principle vary the angle $\beta$ to \emph{any }value between $0$ and $180^{\circ}$. By comparison, such strain range only modifies the angle between the lattice vectors of the honeycomb lattices by, at most, a few degrees \cite{naumis2017electronic}. Superlattice configurations thus offer a rich and accessible experimental platform in which one can vary, or effectively \emph{design}, practically any desired moir\'e geometry. Next, we describe the possible moir\'e geometries under the three types of strain introduced in Section \ref{sec:Strain_types}.

\subsubsection{Uniaxial strain}\label{subsec:uniaxial}

For uniaxial strain, the transformation $\mathbf{F}=\mathbf{T}^{\mathrm{T}}\mathbf{T}$ reads [cf. Eq. \eqref{eq:T}]
\begin{align}
\mathbf{F} & =4\sin^{2}\frac{\theta}{2}\mathbb{I}+\epsilon_u\left(1+\nu\right)\sin\left(\theta\right)\mathrm{R}\left(2\phi\right)\sigma_{x}\nonumber \\
 & \;+\frac{\epsilon_u^{2}}{2}\cos^{2}\frac{\theta}{2}\left[\left(1+\nu^{2}\right)\mathbb{I}+\left(1-\nu^{2}\right)\mathrm{R}\left(2\phi\right)\sigma_{z}\right],\label{eq:F_uniaxial}
\end{align}
where $\sigma_{i}$ are the Pauli matrices. The non-spherical terms in $\mathbf{F}$ (those that can change the moir\'e geometry) depend non-trivially on the interplay between twist and strain. Such interplay is strongly influenced by the Poisson's ratio $\nu$ of the layers. Some general features can still be readily deduced from Eq. \eqref{eq:F_uniaxial}. For instance, Eq. \eqref{eq:F_uniaxial} gives $\mathbf{F}\propto \sin^{2}\left(\theta/2\right)$ if we write $\epsilon_u=\epsilon'\tan\left(\theta/2\right)$. Then one can easily see that Eq. \eqref{eq:cosB} is independent of the twist angle $\theta$, which means that for a fixed $\beta$, the strain magnitude always scale on the twist angle as $\epsilon_u\sim\tan\left(\theta/2\right)$. This again reflects, and noted before, that the lower the twist angle the less strain is needed to modify the geometry of the moir\'e superlattices. 

The transformation given by Eq. \eqref{eq:F_uniaxial} generally yields non-equal length moir\'e vectors with a strain-dependent angle between them. The analysis can be simplify for some special strain configurations. One relevant scenario is with no twist, i.e., when the layers are only subject to uniaxial heterostrain. In that case, the transformation $\mathbf{F}$ reduces to
\begin{equation}
\mathbf{F}=\frac{\epsilon_u^{2}}{2}\left[\left(1+\nu^{2}\right)\mathbb{I}+\left(1-\nu^{2}\right)\mathrm{R}\left(2\phi\right)\sigma_{z}\right]
\end{equation}
While the first term $\propto\left(1+\nu^{2}\right)\mathbb{I}$ is a spherical transformation (so it does not change the moir\'e geometry), the second term $\propto\left(1-\nu^{2}\right)\mathrm{R}\left(2\phi\right)\sigma_{z}$ does change moir\'e geometry according to Eq. \eqref{eq:cosB}. Consequently, the resulting moir\'e pattern is not hexagonal. 

Another relevant strain configuration is that which results in equal length moir\'e vectors. This situation simplifies the analytical analysis, and in particular allows one to obtain what strain parameters result in a particular moir\'e geometry. In terms of the transformation $\mathbf{F}$, the equal length moir\'e vector condition $\mathbf{G}_{1}\cdot\mathbf{G}_{1}=\mathbf{G}_{2}\cdot\mathbf{G}_{2}$ can be stated as
\begin{equation}
\mathbf{F}\left(\mathbf{b}_{1}-\mathbf{b}_{2}\right)\cdot\left(\mathbf{b}_{1}+\mathbf{b}_{2}\right)=0.
\label{F_equal}
\end{equation}
Here we have used that $\mathbf{F}\mathbf{b}_{1}\cdot\mathbf{b}_{2}=\mathbf{F}\mathbf{b}_{2}\cdot\mathbf{b}_{1}$, since $\mathbf{F}$ is a symmetric tensor \cite{gurtin2010mechanics}. Using Eq. \eqref{eq:F_uniaxial} and solving Eq. \eqref{F_equal} for the non-zero strain case gives \cite{escudero2024designing}
\begin{equation}
\epsilon_{eq}=\frac{4}{1-\nu}\cot\left(\frac{\pi}{3}-2\phi\right)\tan\frac{\theta}{2}.
\end{equation}
Note that $\epsilon_{eq}$ is not symmetric under the transformation $\ensuremath{\phi\rightarrow\phi+\pi/3}$ because, and discussed before, the particular moir\'e vectors given by Eq. \eqref{eq:moire_vectors} do not take into account all the symmetrical solutions obtained by appropriately changing the moir\'e vectors construction \cite{escudero2024designing}. The hexagonal symmetry is restored by taking different set of strained reciprocal vectors. For most proposes, it is sufficient to consider $\epsilon_{eq}$ above, and then generalize the results by taking all the missing solutions given by translations $\phi\rightarrow\phi\pm\pi/3$. Keeping that in mind, by solving the angle equation \eqref{eq:cosB} one can find that the strain parameters that yield an angle $\beta$ between the moir\'e vectors read \cite{escudero2024designing}
\begin{align}
\epsilon_{s,r} & =\frac{4s}{1-\nu}\frac{f_{r}}{\sqrt{1-f_{r}^{2}}}\tan\frac{\theta}{2}, \label{eq:strain_equal}
\\
\phi_{s,r} & =-\frac{s}{2}\arccos f_{r}+\frac{\pi}{3}\left(n+\frac{1}{2}\right),
\label{eq:phi_equal}
\end{align}
where
\begin{equation}
f_{r}\left(\nu,\beta\right)=\left(\frac{1-\nu}{1+\nu}\right)\frac{2+\cos\beta+r\sqrt{3}\left|\sin\beta\right|}{1+2\cos\beta}.
\end{equation}
The length of the moir\'e vectors is given by:
\begin{align}
\frac{\left|\mathbf{G}_{i}\right|^{2}}{\left|\mathbf{b}_{i}\right|^{2}} & =\frac{\left(1+\nu\right)^{2}f_{r}^{2}-\left(1-\nu^{2}\right)f_{r}+\left(1-\nu\right)^{2}}{\left(1-f_{r}^{2}\right)\left(1-\nu\right)^{2}}4\sin^{2}\frac{\theta}{2}.
\end{align}
Here $s,r=\pm1$, and $n$ is an integer. The $r=1$ roots correspond to the moir\'e patterns formed through the lateral contraction of the honeycomb lattices, as measured by the Possion's ratio, and thus correspond to larger strain magnitudes. While the $r=-1$ roots are solutions for any angle $\beta$, the roots $r=1$ are only solutions for certain $\beta$. It is interesting to note that the strain angle $\phi$ does not depend on the twist angle $\theta$. This independence of $\phi$ on $\theta$, when $\epsilon=\epsilon_{eq}$, is a direct consequence of our initial assumption that the stress forces act equally but with opposite magnitude in each layer. The twist angle then only influences the needed strain magnitude for an angle $\beta$, as well as the resulting length of the moir\'e vectors. 

\begin{figure*}[t]
\centering
    \includegraphics[width=\linewidth]{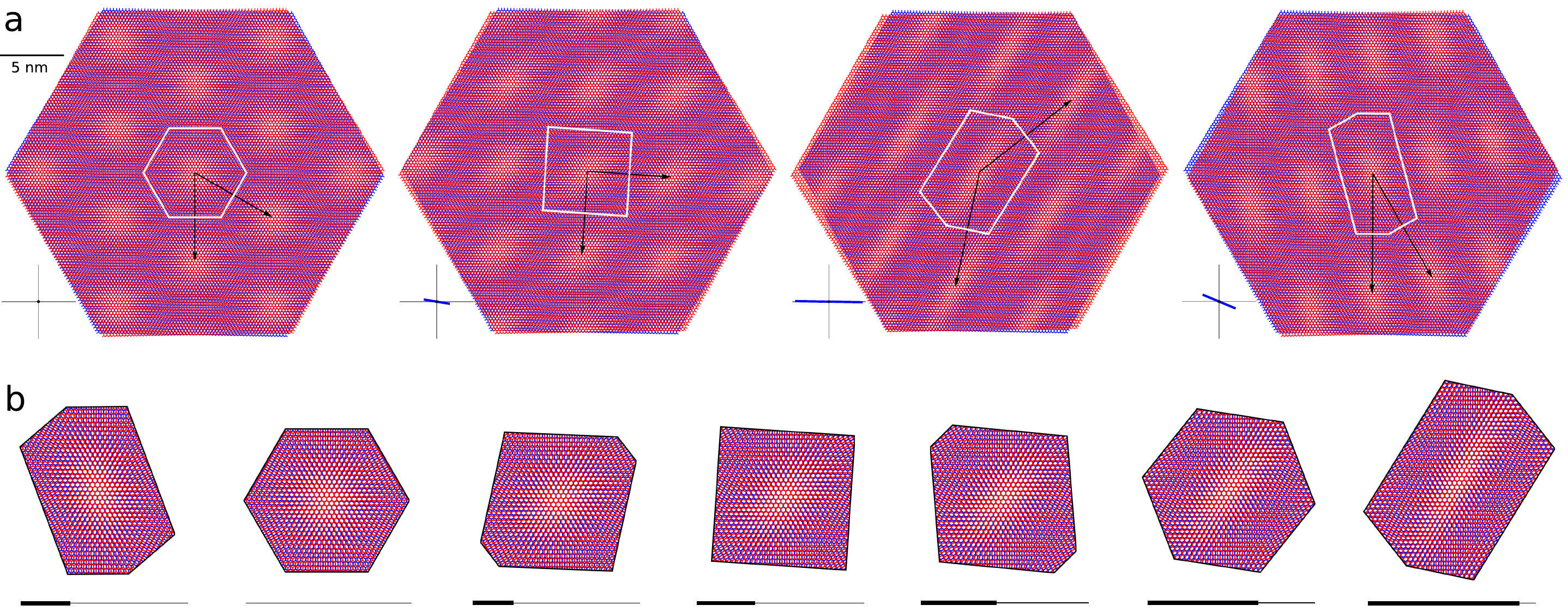}
    \caption{Examples of moir\'e patterns in two equal hexagonal layers with relative twist $\theta=2^{\circ}$ and uniaxial heterostrain. All cases shown correspond to equal length moir\'e vectors (Section \ref{subsec:uniaxial}). Panel (a) shows three moir\'e pattern corresponding to (from left to right): $\epsilon=0$ (no strain), $\epsilon\approx1.64\%$, $\phi\approx-9.4^{\circ}$, $\epsilon\approx4.44\%$, $\phi\approx-0.9^{\circ}$ and $\epsilon\approx-2.28\%$, $\phi\approx-22.7^{\circ}$. The relative angles between the real space moir\'e vectors are, respectively, $\beta_{R}=60^{\circ},90^{\circ},140^{\circ},30^{\circ}$. In each case the Wiger-Seitz cell of the superlattice is shown in white. The small inset below each panel show the strain magnitude (in a scale of $5\%$) and the strain direction relative to the $x$ axis. Panel (b) shows the evolution of the Wigner-Seitz cell and the \emph{repeated }moir\'e pattern within it, for moir\'e vector angles (from left to right) $\beta_{R}=40^{\circ},60^{\circ},80^{\circ},90^{\circ},100^{\circ},120^{\circ},140^{\circ}$. For each case, the bar underneath indicates the strain magnitude (thick line) in a scale of $5\%$ (thin line). Adapted under the terms of the CC BY license from Ref. \cite{escudero2024designing}. Copyright (2024) by the American Physical Society.}
\label{fig:Moire_Patterns}
\end{figure*}

Some moir\'e patterns obtained from uniaxial strain can be seen in Fig. \ref{fig:Moire_Patterns}. Due to the magnifying effect of the moir\'e, small changes in the strain parameters can lead to significant changes in the resulting superlattice geometry. Consequently, it becomes imperative to have a precise control of the strain parameters in order to design particular moir\'e geometries \cite{pena2023moire} (see Section \ref{sec:Strain_Implementation}). Conversely, such strong dependence on the strain parameters indirectly explains the plethora of different moir\'e geometries seen in experiments \cite{alden2013strain, bai2020excitons, hesp2024cryogenic}, particularly in those where strain effects unavoidable arise during the superlattice assemble \cite{kazmierczak2021strain}. 

The observed shape of the strain moir\'e patters in Fig. \ref{fig:Moire_Patterns} reveals several noteworthy features. For instance, we see that the modification of the moir\'e geometry by the strain leads to significant changes in the corresponding Wigner--Seitz cell. By construction, such cell reflects the new symmetries of the AA stacking regimes. The change in the moir\'e periodicity is accompanied by a stretch of the AA stacking within the primitive cell, which gives a direct reflection of the strain magnitude (i.e., the larger the stretch, the larger the strain). A complete description of the strained moir\'e patterns must thus take into account both the geometry of the moir\'e vectors, and the shape of the stretched AA regions. The latter naturally accounts, in a picture manner, the atomic basis of both layers within the supercell.

\subsubsection{Shear strain}

For shear strain the transformation $\mathbf{F}$ reads
\begin{equation}
\mathbf{F}=\left(4\sin^{2}\frac{\theta}{2}+\epsilon_{s}^{2}\cos^{2}\frac{\theta}{2}\right)\mathbb{I}-2\epsilon_{s}\sin\left(\theta\right)\mathrm{R}\left(2\varphi\right)\sigma_{z}.\label{eq:F_shear} 
\end{equation}
The last term, which depends on the interplay between twist and strain, is not a spherical tensor so it changes the moir\'e geometry. In other words, the moir\'e patterns with both twist and shear strain are not hexagonal \cite{kogl2023moire, escudero2024designing}. The possible strained configurations are similar to those described in the previous section for uniaxial heterostrain (this analogy becomes exact with a combination of shear and biaxial strain, see Section \ref{sec:Shear_and_Biaxial}). 

\subsubsection{Biaxial strain}\label{sec:Biaxial}

For biaxial strain the transformation $\mathbf{F}$ reads
\begin{equation}
\mathbf{F}=\left(4\sin^{2}\frac{\theta}{2}+\epsilon_{b}^{2}\cos^{2}\frac{\theta}{2}\right)\mathbb{I}.
\label{eq:F_biaxial}
\end{equation}
As this is, as expected, a spherical tensor, a biaxial strain does not change the moir\'e geometry. Its effect is to only modify the orientation and length of the moir\'e vectors. However, both changes are important and can actually significantly influence the geometric and electronic properties of the system \cite{escudero2024designing}. 

The prevalence of a hexagonal moir\'e pattern in the presence of biaxial strain implies that there are combinations of $\epsilon_{b}$ and $\theta$ that result in the same moir\'e periodicity as with only a twist angle $\theta_{eq}$. Indeed, by comparing Eq. \eqref{eq:F_biaxial} with the non-strain case $\mathbf{F}=4\sin^{2}\left(\theta_{eq}/2\right)\mathbb{I}$ [cf. Eq. \eqref{eq:F_0}], one readily finds that this occurs when
\begin{equation}
\sin^{2}\left(\theta/2\right)=\frac{\sin^{2}\left(\theta_{eq}/2\right)-\epsilon_{b}^{2}/4}{1-\epsilon_{b}^{2}/4}.
\label{eq:periodicity_biaxial}
\end{equation}
The moir\'e periodicity for a given twist angle $\theta$ decreases as the strain increases. The above condition does not, however, guarantee that both moir\'e patterns (with and without strain) would be aligned because their orientation may differ. This can be important in superlattice configurations made up of three (or more) layers with a lattice mismatch, and where only two layers are relatively rotated. In general, the angle $\alpha$ of the moir\'e pattern orientation, with respect to the non-strain case, reads \cite{escudero2024designing}
\begin{equation}
\cos\alpha=\frac{1}{\sqrt{1+\epsilon_{b}^{2}\cot^{2}\left(\theta/2\right)/4}}.
\label{eq:cos_biaxial}
\end{equation}
The change of orientation depends on the interplay between twist and the biaxial strain: it only occurs in the presence of both.

A relevant example of the importance of matching the periodicity and orientation condition occurs in heterostructures of TBG/hBN, in which hBN acts as a substrate of TBG \cite{yankowitz2012emergence, xue2011scanning, decker2011local}. This situation can be described by considering three equal graphene monolayers, but with a biaxial strain of magnitude $\ensuremath{\epsilon_{b}\sim1-a_{g}/a_{h}=0.016}$ in the bottom layer \cite{cea2020band}, in order to account for the lattice mismatch between graphene ($a_{g}=2.46\;\textrm{Å}$) and hBN ($a_{h}=2.50\;\textrm{Å}$). Since the moir\'e pattern remains hexagonal with twist and biaxial strain, the orientation match between the graphene/hBN moir\'e and the TBG moir\'e occurs when the angle $\alpha$ in Eq. \eqref{eq:cos_biaxial} is $60^{\circ}$ (or integers values of it), so that $\cos\alpha=\pm1/2$. Combining this with the equal periodicity condition of Eq. \eqref{eq:periodicity_biaxial}, one finds the moir\'e alignment conditions \cite{long2022atomistic, long2023electronic, cea2020band, escudero2024designing}
\begin{align}
\theta_{B} & \approx\frac{\epsilon_{b}}{\sqrt{3}}\sim0.53^{\circ},\\
\theta_{T} & \approx\sqrt{\theta_{B}^{2}+\epsilon_{b}^{2}}\sim1.06^{\circ},
\end{align}
where $\theta_{B}$ is the twist angle between hBN and the graphene layer on top, and $\theta_{T}$ is twist angle between the graphene monolayers. 

\subsubsection{Shear and biaxial strain}\label{sec:Shear_and_Biaxial}

In the presence of both shear and biaxial strains, the lattices change both their shape and size. The corresponding strain tensor is given by
\begin{equation}
\mathcal{E}=\left(\begin{array}{cc}
\epsilon_{b}-\epsilon_{s}\sin2\varphi & \epsilon_{s}\cos2\varphi\\
\epsilon_{s}\cos2\varphi & \epsilon_{b}+\epsilon_{s}\sin2\varphi
\end{array}\right).
\end{equation}
As noted above, any strain tensor can be generally decomposed as a mixture of biaxial and shear strains, cf. Eq. \eqref{eq:strain_general}. The corresponding transformation $\mathbf{F}$ is given by
\begin{align}
\mathbf{F} &=\left[4\sin^{2}\left(\theta/2\right)+\left(\epsilon_{s}^{2}+\epsilon_{b}^{2}\right)\cos\left(\theta/2\right)\right]\mathbb{I}\nonumber\\
  &\;-2\epsilon_{s}R\left(2\varphi\right)\left[\sin\left(\theta\right)\sigma_{z}-\epsilon_{b}\cos^{2}\left(\theta/2\right)\sigma_{x}\right].
\label{eq:F_shear_biaxial} 
\end{align}
The last, shear-dependent term is non-spherical and thus modifies the moir\'e geometry. Note that $\mathbf{F}$ above is given by the sum of the corresponding transformations for shear and biaxial strain, given by Eqs. \eqref{eq:F_shear} and \eqref{eq:F_biaxial}, plus a non-spherical term $\sim\epsilon_{s}\epsilon_{b}R\left(2\varphi\right)\cos^{2}\left(\theta/2\right)\sigma_{x}$
that depends on the interplay between the twist angle and both types of strain. Thus the change in the geometry of the moir\'e vectors is not given, in general, by just its change due to the effect of biaxial and shear strain separately. 

The possible geometries of the strained moir\'e pattern, as given by the transformation \eqref{eq:F_shear_biaxial}, are analogous to those that result from uniaxial heterostrain. One can actually state this analogy precisely by noting that both cases are equivalent by the correspondence \cite{escudero2024designing}
\begin{align}
\epsilon & \rightarrow\epsilon_{s}+\epsilon_{b},\nonumber \\
\nu & \rightarrow\frac{\epsilon_{s}-\epsilon_{b}}{\epsilon_{s}+\epsilon_{b}},\label{eq:correspondence}\\
\phi & \rightarrow\varphi+\pi/4.\nonumber 
\end{align}
This allows one to readily obtain the geometrical properties of the strained moir\'e patterns due to shear and biaxial strain from those previously analyzed for uniaxial heterostrain (Section \ref{subsec:uniaxial}). In particular, one can directly replace the above correspondence in Eqs. \eqref{eq:strain_equal} and \eqref{eq:phi_equal} and obtain the needed combination of shear and biaxial strain in order to obtain equal-length moir\'e vectors with an angle $\beta$ between them. It should be noted that this analogy between different kinds of strain is exact, it does not just alludes to an equivalence of periodicity or orientation. Indeed, the above correspondence leads to equivalent uniaxial and shear+biaxial strain tensors, so that each lattice is exactly strained in the same way. Thus, one can think of uniaxial heterostrain as a particular combination of biaxial and shear strain \cite{kogl2023moire, escudero2024designing}. As the Poisson's ratio is generally fixed by the properties of the material, this analogy necessarily fixes the ratio between shear and biaxial strain magnitude
\begin{equation}
\frac{\epsilon_{b}}{\epsilon_{s}}\rightarrow\frac{1-\nu}{1+\nu}.
\end{equation}
Conversely, a combination of shear and biaxial strain can be considered as uniaxial heterostrain with an effective Poisson's ratio that depends on $\epsilon_{b}/\epsilon_{s}$. In this sense, a combination of shear and biaxial strain offers a wider range of strain configurations from which one can design different moir\'e patterns. 

\begin{figure*}[t]
\centering
    \includegraphics[width=0.9\linewidth]{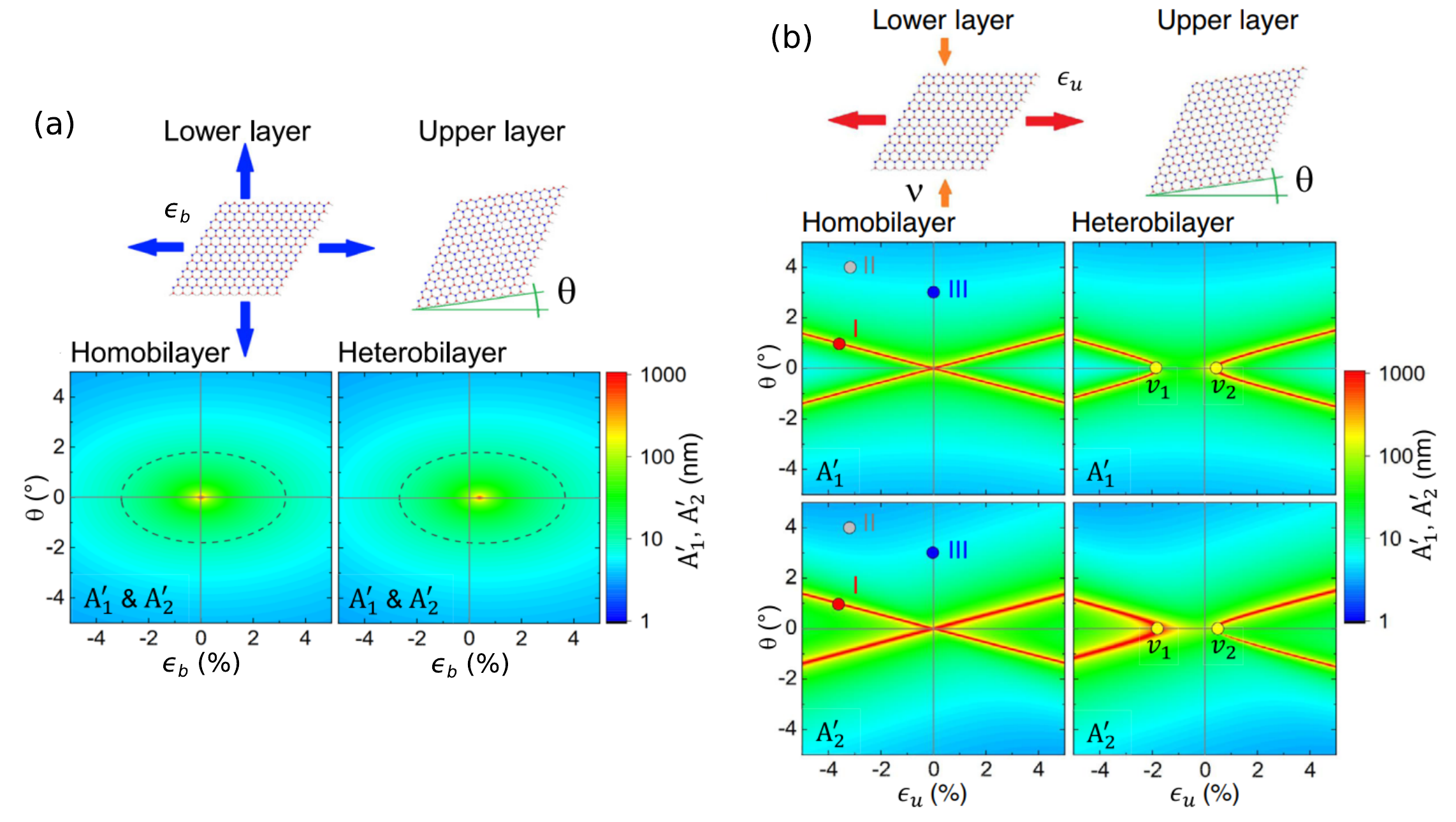}
    \caption{Variation of the real space moir\'e vectors lengths $A'_{1}=\left|\mathbf{G}_{1}^{R}\right|$ and $A'_{2}=\left|\mathbf{G}_{2}^{R}\right|$, as a function of the twist angle and the strain magnitude, for the cases of two homobilayer ($\mathrm{WSe_{2}-WSe_{2}}$) and heterobilayer ($\mathrm{WSe_{2}}-\mathrm{MoSe_{2}}$) hexagonal lattices. The configuration is such that the top layers is only twisted and the bottom layer is only strained. Panel (a) shows the moir\'e lengths $A'_{1},A'_{2}$ (equal) under biaxial strain magnitude $\epsilon_{b}$; dotted ellipses indicate the regions of moir\'e lenghts $A'_{1},A'_{2}=10\,\mathrm{nm}.$ Panel (b) shows the moir\'e lengths $A'_{1}$ (top) and $A'_{2}$ (bottom) under uniaxial strain magnitude $\epsilon_{u}$ at fixed direction $\phi=0^{\circ}$. The spots labeled as I, II and III indicate the unidimensional, square and hexagonal moir\'e geometries, respectively (cf. Section \ref{sec:Moire_special}). The points $\nu_{1}$ and $\nu_{2}$ in the heterobilayer case indicate the vertexes of the divergence curves (unidimensional patterns). Adapted under the terms of the CC BY license from Ref. \cite{kogl2023moire}. Copyright (2023) The author(s).}
\label{fig:Strain_Homobilayer_Heterobilayer}
\end{figure*}

\subsection{Hexagonal heterobilayers}\label{sec:heterobilayers}

The analysis of the previous section can be directly generalized to the case of two hexagonal layers with different lattice constant. We refer to these as \textit{heterobilayers}. Since then the non-deformed lattice vectors in each lattice are different, one cannot determine the moir\'e vectors through a transformation $\mathbf{T}$ as in Eq. \eqref{eq:T}. However, one can still define a generalized transformation $\mathbf{T}_{h}$ by accounting the lattice mismatch with a biaxial strain $\mathcal{E}_{b}=\mathbb{I}\epsilon_{b}$ \cite{cosma2014moire, kogl2023moire}. Since the moir\'e pattern only depends on relative deformations, it is convenient to introduce a biaxial strain tensor $\pm\mathcal{E}_{b}/2$ so that the moir\'e vectors in heterobilayers can be determined as $\mathbf{G}_{i}=\mathbf{T}_{h}\mathbf{b}_{i}$, where
\begin{equation}
\mathbf{T}_{h}=\sum_{\ell=\pm}\ell\left(\mathbb{I}+\mathcal{E}_{\ell}\right)\mathrm{R}\left(-\ell\theta/2\right)\left(\mathbb{I}+\ell\mathcal{E}_{b}/2\right).
\label{eq:Th}
\end{equation}
The analysis of Section \ref{sec:homobilayers} can then be directly generalized to the heterobilayer case by simply replacing $\mathbf{T}$ with $\mathbf{T}_{h}$.

Figure \ref{fig:Strain_Homobilayer_Heterobilayer} shows the variation in the length of the real space moir\'e vectors $A'_{1},A'_{2}=\left|\mathbf{G}_{1}^{R}\right|,\left|\mathbf{G}_{2}^{R}\right|$, as a function of the twist angle and the strain magnitude (biaxial or uniaxial), for the cases of hexagonal homobilayers and heterobilayers. The results correspond to $\mathrm{WSe_{2}}$ and $\mathrm{MoSe_{2}-WSe_{2}}$ homobilayers and heterobilayers, with lattice constant $a_{\mathrm{MoSe_{2}}}=0.329\,\mathrm{nm}$ and $0.4\%$ smaller for $a_{\mathrm{WSe_{2}}}$, and Poisson's ratios $\nu_{\mathrm{MoSe_{2}}}=0.23$, $\nu_{\mathrm{WSe_{2}}}=0.19$. The results correspond to a particular non-symmetric configuration in which the upper layer is only twisted while the lower layer is only strained.

In the biaxial strain case, the moir\'e length diverges when $\theta=0$ due to the absence of a moir\'e pattern (infinite moir\'e length); for homobilayers this occurs when $\epsilon_{c}=0$, but for heterobilayers this occurs when $\epsilon_{c}\sim0.4\%$ due to the lattice mismatch. For the uniaxial strain case, the change in the moir\'e lengths is radically different, as the moir\'e lengths now diverge along straight-lines curves in the twist-strain plane. These divergences represent the formation of quasi one-dimensional channels that follow a linear relation between $\theta$ and $\epsilon$ [see Section \ref{sec:1D_channels} and Eq. \eqref{eq:1D_uniaxial}]. Furthermore, for uniaxial strain the moir\'e lengths $A'_{1}$ and $A'_{2}$ are not, in general, equivalent. The main difference between the homobilayer and the heterobilayer case, for uniaxial strain, is revealed by the no-crossing of the moir\'e length divergences from negative to positive strain. The two vertexes $\nu_{1}$ and $\nu_{2}$ of the divergences are determined by the lattice mismatch, becoming equal ($\nu_{1}=\nu_{2}$) when the layers are the same (homobilayer case). 

\begin{figure*}[t]
\centering
    \includegraphics[width=0.9\linewidth]{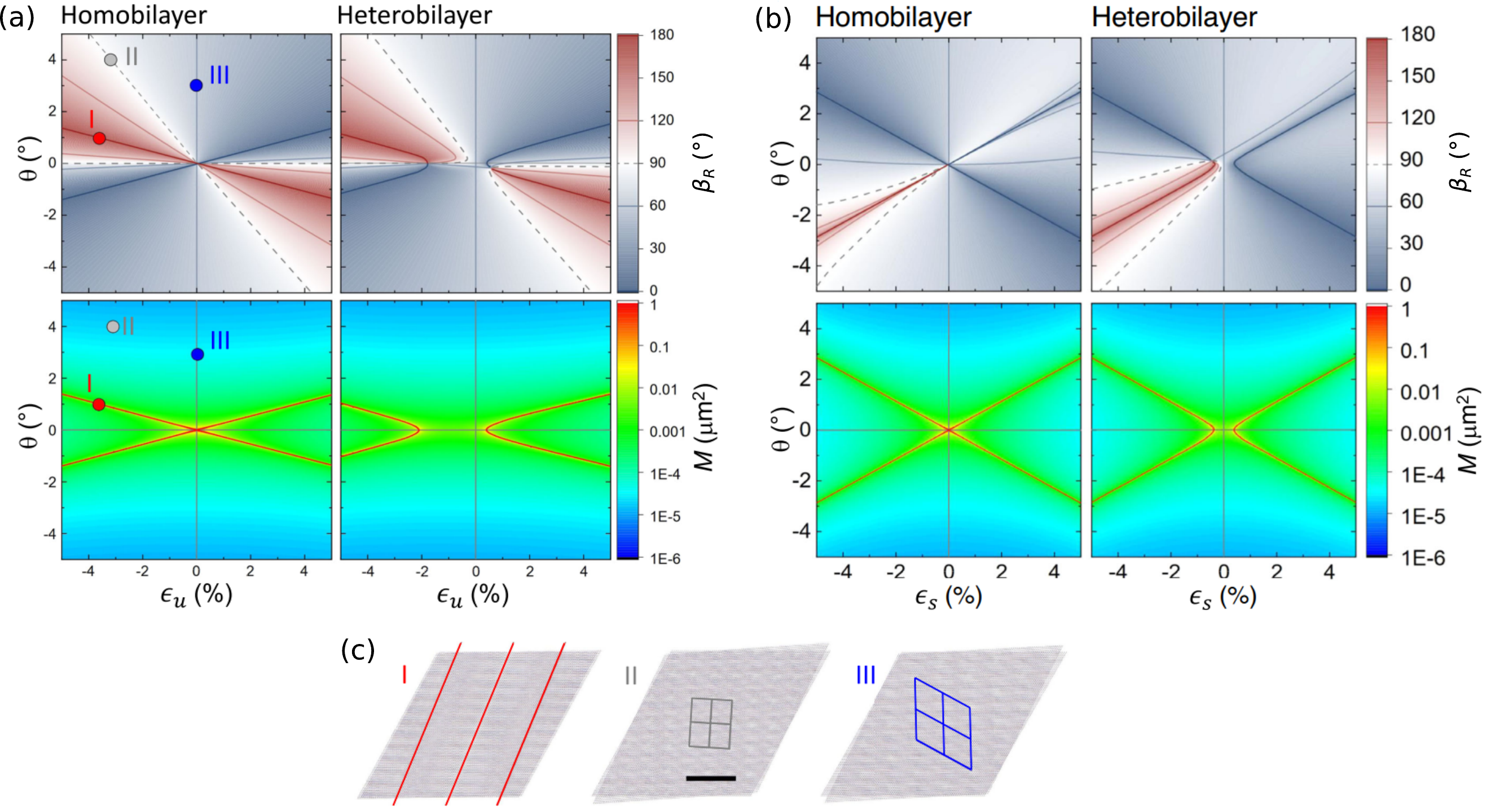}
    \caption{Variation of the real-space angle $\beta_{R}$ between the moir\'e vectors (top) and the moir\'e area $M=\left|\mathbf{G}_{1}^{R}\times\mathbf{G}_{2}^{R}\right|$ (bottom), as a function of the twist angle and the strain magnitude, for the cases of two homobilayer ($\mathrm{WSe_{2}-WSe_{2}}$) and heterobilayer ($\mathrm{WSe_{2}}-\mathrm{MoSe_{2}}$) hexagonal lattices (same configuration as in Figure \ref{fig:Strain_Homobilayer_Heterobilayer}). Panel (a) shows the case of uniaxial strain with magnitude $\epsilon_{u}$ at fixed direction $\phi=0^{\circ}$. Panel (b) shows the case of shear strain with magnitude $\epsilon_{s}$ at fixed direction $\phi=0^{\circ}$. In both cases, the spots labeled as I, II and III indicate the unidimensional, square and hexagonal moir\'e geometries (see also Section \ref{sec:Moire_special}). Adapted under the terms of the CC BY license from Ref. \cite{kogl2023moire}. Copyright (2023) The author(s).}
\label{fig:Moire_Uniaxial_Shear}
\end{figure*}

Figure \ref{fig:Moire_Uniaxial_Shear} shows the angle $\beta_{R}$ between the moir\'e vectors and the area $M=\left|\mathbf{G}_{1}^{R}\times\mathbf{G}_{2}^{R}\right|$ of the moir\'e unit cell, as a function of twist and uniaxial or shear strain magnitudes, again for the homobilayer and heterobilayer cases (same configuration as in Figure \ref{fig:Strain_Homobilayer_Heterobilayer}). The angle variation $\beta_{R}$ represents the change in the moir\'e geometry due to twist and uniaxial strain (section \ref{subsec:uniaxial}). In both strain cases the angle $\beta_{R}$ can effectively vary from $0$ to $180$ degrees even for relatively small strain magnitudes, specially at small twist angles. The labels I, II and III identify, respectively, three special patterns: unidimensional (I, $\beta_{R}=0^{\circ}$ or $\beta_{R}=180^{\circ}$), square (II, $\beta_{R}=90^{\circ}$) and triangular (III, $\beta_{R}=60^{\circ}$), shown in the bottom panels (see Section \ref{sec:Moire_special} for more details). The formation of the unidimensional channels is reflected in the divergence of the moir\'e lengths. As in Figure \ref{fig:Strain_Homobilayer_Heterobilayer}, when comparing the homobilayer to the heterobilayer case, the main difference is that the curves of particular moir\'e geometries (I, II and III) do not intercept. For heterobilayers there is, in general, a twist and strain threshold to obtain certain moir\'e geometries, due to the lattice mismatch.

\subsection{Monoclinic lattices}

Let us now present the generalized geometric description of moir\'e patterns arising for any two arbitrary monoclinic lattices. We will mostly follow the formalism of Ref. \cite{kogl2023moire}. The lattices vector of the top and bottom layers can be generally written as
\begin{align}
\mathbf{a}_{i,+} & =a_{i}\left(1+\delta\right)R\left(\psi_{i}+\theta\right)\left(\begin{array}{c}
1\\
0
\end{array}\right), \label{eq:ai_mono}\\
\mathbf{a}_{i,-} & =a_{i}R\left(\psi_{i}\right)\left(\begin{array}{c}
1\\
0\label{eq:ab_mono}
\end{array}\right),
\end{align}
where $a_{i}$ are the lattice constants and $\theta$ is the usual twist angle between the layers. The angle $\psi_{i}$ is taken as
\begin{equation}
\psi_{i}=\begin{cases}
\theta_{0} & i=1\\
\theta_{0}+\alpha & i=2
\end{cases},
\end{equation}
where $\theta_{0}$ is the overall rotation of the lattices with respect to the $x$-axis, and $\alpha$ is the angle between the primitive vectors of each layer. The factor $\delta$ in Eq. \eqref{eq:ai_mono} accounts for different lattice constant in each layer. 

Upon introducing strain into the system, the new lattice vectors $\mathbf{a}'_{i,\pm}$ transform as
\begin{align}
\mathbf{a}'_{i,+} & =\left(\mathbb{I}+\mu\mathcal{E}\right)\mathbf{a}_{i,+},\\
\mathbf{a}'_{i,-} & =\left(\mathbb{I}+\mathcal{E}\right)\mathbf{a}_{i,-}.
\end{align}
Here the parameter $0<\mu<1$ accounts for different strain magnitudes in each layer. The heterostrain scenario corresponds to $\mu=0$ (only strain in one layer). The opposite case $\mu=1$ is referred as homostrain (equal strain in both layers). In general the value of $\mu$ can be set to match different experimental setups. 

The real space moir\'e vectors $\mathbf{G}_{i}^{R}$ can be computed analytically, with the result \cite{kogl2023moire}
\begin{equation}
\mathbf{G}_{i}^{R}=\frac{a_{i}\left(1+\delta\right)}{\Delta}X\left(\mathcal{E}\right)R\left(\psi_{i}\right)\left(\begin{array}{c}
1\\
0
\end{array}\right),
\end{equation}
where 
\begin{align}
X\left(\mathcal{E}\right) & =\left(1+\delta\right)c_{\mu}\left[\left(1+\epsilon_{b}\right)\mathbb{I}+\epsilon_{s}S\left(\phi\right)\right]\nonumber \\
 & \;-c_{1}\left[\left(1+\mu\epsilon_{b}\right)R\left(\theta\right)+\mu\epsilon_{s}S\left(\phi+\theta\right)\right] ,\nonumber\\
\Delta & =c_{\mu}\left(1+\delta\right)^{2}+c_{1}\nonumber \\
 & -2\left(1+\delta\right)\left[\left(1+\epsilon_{b}+\mu\epsilon_{b}\right)+\mu\left(\epsilon_{b}^{2}-\epsilon_{s}^{2}\right)\right]\cos\theta,\nonumber\\
c_{1} & =\left(1+\epsilon_{b}\right)^{2}-\epsilon_{s}^{2},\nonumber\\
c_{\mu} & =\left(1+\mu\epsilon_{b}\right)^{2}-\mu^{2}\epsilon_{s}^{2}.
\end{align}
Here the strain parameters $\epsilon_{b}$, $\epsilon_{s}$, the shear matrix $S\left(\phi\right)$ and the angle $\phi$ are given by Eqs. \eqref{eq:strain_equiv} and \eqref{eq:shear_matrix}. The area of the moir\'e unit cell follows directly as
\begin{equation}
\left|\mathbf{G}_{1}^{R}\times\mathbf{G}_{2}^{R}\right|=a_{1}a_{2}\left|\frac{\sin\alpha}{\Delta}\right|\left(1+\delta\right)^{2}c_{1}c_{\mu}.
\end{equation}
The angle $\beta_{R}$ between the real-space moir\'e vectors $\mathbf{G}_{1}^{R}$ and $\mathbf{G}_{2}^{R}$ can be calculated as in Eq. \eqref{eq:cosB}. Alternatively, one can compute the angle through the relation \cite{kogl2023moire}
\begin{equation}
\sin\beta_{R}=\left|\hat{\mathbf{G}}_{1}^{R}\times\hat{\mathbf{G}}_{2}^{R}\right|,
\end{equation}
where $\hat{\mathbf{G}}_{i}^{R}=\mathbf{G}_{i}^{R}/\left|\mathbf{G}_{i}^{R}\right|$. These formulas are general and hold for any kind of monoclinic lattices and strain (within the limit of small deformations). The particular hexagonal cases follow by taking $\alpha=60^{\circ}$ with $\delta=0$ (homobilayer, Section \ref{sec:homobilayers}) and $\delta\neq0$ (heterobilayer, Section \ref{sec:heterobilayers}). The initial assumption that only the top layer is twisted can be relaxed by setting $\psi_{i}\rightarrow\psi_{i}+\theta_{-}$ in Eq. \eqref{eq:ab_mono}, where $\theta_{-}$ is now twist in the bottom layer. The particular symmetric twist configuration $\pm\theta/2$ follows by taking $\psi_{i}\rightarrow\psi_{i}\mp\theta/2$ in Eqs. \eqref{eq:ai_mono} and \eqref{eq:ab_mono}, and in all subsequent expressions for the moiré vectors.

\subsection{Special moir\'e patterns}\label{sec:Moire_special}

The formalism presented in the previous sections implies that the interplay between twist and strain in stacked two-dimensional materials can lead to many different moiré geometries. Crucially, due to the magnifying effect of the moiré, this plethora of moiré geometries can be archived even with very small strain magnitudes, well within the experimental regime (see Section \ref{sec:Strain_Implementation} and Table \ref{table1}). Thus, with the right combination of twist and strain it becomes possible to engineer \emph{any} particular 2D moiré geometry.

In this section, we use the general formalism of the Sections III.A-C to obtain and describe three relevant moiré geometries that can be realized with twist and strain: (1) One-dimensional moiré patterns, (2) Square moiré patterns, and (3) Hexagonal moiré patterns.

\subsubsection{Quasi-unidimensional patterns}\label{sec:1D_channels}

The significant deformation of the moir\'e geometry under strain can lead to a critical situation in which the moir\'e vectors become collinear \cite{cosma2014moire, sinner2023strain, kogl2023moire, escudero2024designing}. This result in the formation of quasi-unidimensional moir\'e channels that have been seen in numerous experiments (see Section \ref{sec:Strain_Implementation}). In reciprocal space, as the critical limit is approached the moir\'e Brillouin zone is squeezed, up until it collapses at a certain critical strain value. At such critical point, the moir\'e vector become collinear, and thus are no longer independent. This occurs when the transformation $\mathbf{T}$, which determines the moir\'e vectors through $\mathbf{g}_{i}=\mathbf{T}\mathbf{b}_{i}$ [Eq. \eqref{eq:T}], is such that \cite{sinner2023strain}
\begin{equation}
\mathbf{T}\left(\mathbf{b}_{1}-\frac{\alpha_{1}}{\alpha_{2}}\mathbf{b}_{2}\right)=\mathbf{0},
\end{equation}
where $\alpha_{1}$ and $\alpha_{2}$ are real number. Since the reciprocal lattice vectors $\mathbf{b}_{i}$ are linearly independent, the above equation is satisfied only if
\begin{equation}
\mathrm{det}\,\mathbf{T}=0.
\end{equation}
This result is independent of the reciprocal vectors, and thus valid for any type of lattice. 

\begin{figure*}[t]
\centering
    \includegraphics[width=\linewidth]{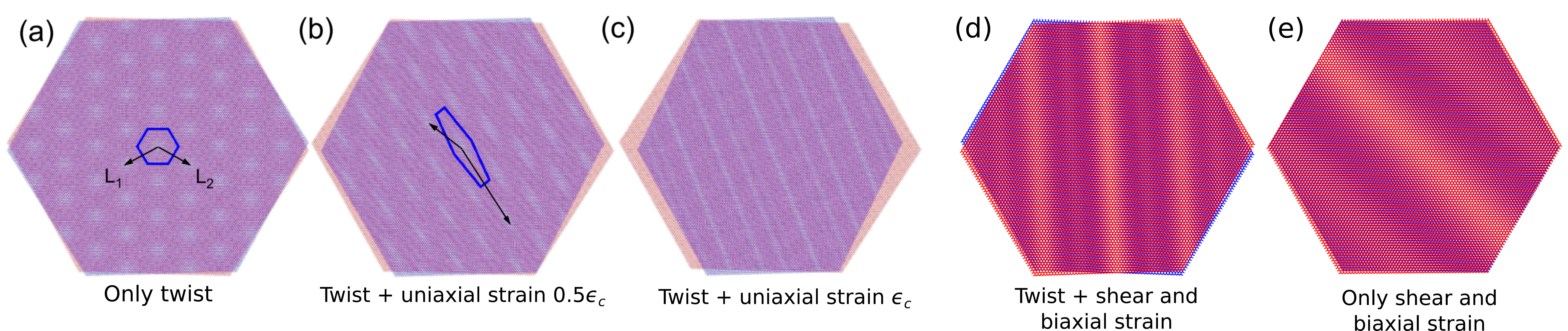}
    \caption{Formation of one-dimensional moir\'e patterns in twisted and strained bilayer graphene. Panels (a) to (c) correspond to a relative twist $\theta=3^{\circ}$ and uniaxial strain with direction $\phi=0^{\circ}$ and magnitudes $\epsilon=0$ , $\epsilon=0.5\epsilon_{c}$ and $\epsilon=\epsilon_{c}$, respectively, where $\epsilon_{c}$ is the critical strain magnitude given by Eq. \eqref{eq:1D_uniaxial}. Reprinted figure with permission from Ref. \cite{sinner2023strain}. Copyright (2023) by the American Physical Society. Last two panels correspond to: (d) A relative twist $\theta=2^{\circ}$ with biaxial strain $\epsilon_{b}=0.5\%$ and shear strain $\epsilon_{s}=\sqrt{\theta^{2}+\epsilon_{b}^{2}}\sim3.5\%$ [cf. Eq. \eqref{eq:1D_channel_2}]; (e) only shear and biaxial strain with magnitudes $\epsilon_{b}=\epsilon_{s}=1.5\%$ [cf. Eq. \eqref{eq:1D_channel_3}]. Adapted under the terms of the CC BY license from Ref. \cite{escudero2024designing}. Copyright (2024) by the American Physical Society.}
\label{fig:Moire_1D}
\end{figure*}

In the particular case of uniaxial heterostrain, the collapse condition $\mathrm{det}\,\mathbf{T}=0$ gives a critical strain magnitude \cite{cosma2014moire, sinner2023strain}
\begin{equation}
\epsilon_{c}=\pm\frac{2}{\sqrt{\nu}}\tan\frac{\theta}{2},
\label{eq:1D_uniaxial}
\end{equation}
independent of the strain angle $\phi$. At low twist angles the critical strain reduces to $\epsilon_{c}\sim\pm\theta/\sqrt{\nu}$ (for graphene $\epsilon_{c}\sim\pm5\theta/2$). Thus the required critical strain magnitude is relatively small at low twist angles, well within the experimental range. Equation \eqref{eq:1D_uniaxial} describes with impressive accuracy the formation of unidimensional channels reported in numerous experiments \cite{mendoza2021strain, bai2020excitons}. 

The critical condition can also take place under other types of strain configurations. For instance, by the correspondence given by Eq. \eqref{eq:correspondence} one readily obtains that unidimensional channels can also arise under shear and biaxial strain if \cite{escudero2024designing}
\begin{equation}
\sqrt{\epsilon_{s}^{2}-\epsilon_{b}^{2}}=\pm2\tan\frac{\theta}{2},
\label{eq:1D_channel_2}
\end{equation}
independently of the shear angle $\varphi$. The above result also follows from the critical condition $\mathrm{det}\,\mathbf{F}=0$ using Eq. \eqref{eq:F_shear_biaxial} (since $\mathrm{det}\,\mathbf{F}=\mathrm{det}\,\mathbf{T}^{\mathrm{T}}\mathrm{det}\,\mathbf{T}$). From it one can further deduce special but relevant cases, such as the formation of 1D-channels solely by twist and shear strain, which occurs if
\begin{equation}
\epsilon_{s}=\pm2\tan\frac{\theta}{2}.
\end{equation}
This critical shear strain is $\sim\sqrt{\nu}\approx0.4$ times smaller than the one required for uniaxial heterostrain. Equation \eqref{eq:1D_channel_2} also predicts the possibility of unidimensional channels arising solely by strain (i.e., $\theta=0$), which occurs when \cite{escudero2024designing}
\begin{equation}
\epsilon_{s}=\pm\epsilon_{b}.
\label{eq:1D_channel_3}
\end{equation}
This is a remarkable result: no matter the magnitude of the strains, as long as they are equal in magnitude the result are collinear moir\'e vectors. The shear angle $\varphi$ only modifies the orientation and length of the collinear moir\'e vectors. Figure \ref{fig:Moire_1D} shows three examples of one-dimensional moir\'e patterns arising from: (i) Twist and uniaxial strain; (ii) Twist and biaxial and shear strain; and (iii) Only biaxial and shear strain.

The critical condition given by Eq. \eqref{eq:1D_uniaxial}, or by Eq. \eqref{eq:1D_channel_2}, relies on the common assumption that the strain forces act with equal magnitude but in opposite direction in each layer, i.e., $\mathcal{E}_{\pm}=\pm\mathcal{E}/2$ (symmetric configuration). Relaxing this condition gives a wider range of strain configurations that can lead to the formation of unidimensional channels. For instance, for homobilayers without a twist, the transformation given by Eq. \eqref{eq:T} generalizes to $\mathbf{T}=\mathcal{E}_{-}-\mathcal{E}_{+}$. A combination of biaxial and shear strain (with shear angle $\varphi=0$), then satisfies the critical condition $\mathrm{det}\,\mathbf{T} =0$ when
\begin{equation}
\epsilon_{s,+}-\epsilon_{s,-}=\pm\left(\epsilon_{b,+}-\epsilon_{b,-}\right),
\end{equation}
where $\epsilon_{s,\pm}$ and $\epsilon_{b,\pm}$ are the shear and biaxial strain magnitude in each monolayer. The above can be satisfied even if only layer is strained. The generalization to arbitrary shear angles in each lattice further increases the possible only-strain configurations that yield unidimensional channels.

Insights into the moir\'e Brillouin zone collapse can be obtained by studying the behavior of the moir\'e vectors as the critical strain limit is approached. In the particular case of one-dimensional channels arising from uniaxial heterostrain along the direction $\phi=0$, for a strain magnitude $\epsilon=x\epsilon_{c}$ with $0<x<1$, the real space moir\'e vectors $\mathbf{L}_{i}$ can be recasted as \cite{sinner2023strain}
\begin{align}
\mathbf{L}_{1} & =\frac{\sqrt{3}a}{4\sin\left(\theta/2\right)}\frac{1}{1-x^{2}}\left(\begin{array}{c}
-\sqrt{3}+x\sqrt{\nu}\\
-1+x\sqrt{3/\nu}
\end{array}\right),\nonumber\\
\mathbf{L}_{2} & =\frac{\sqrt{3}a}{4\sin\left(\theta/2\right)}\frac{1}{1-x^{2}}\left(\begin{array}{c}
\sqrt{3}+x\sqrt{\nu}\\
-1-x\sqrt{3/\nu}
\end{array}\right).
\end{align}
The moir\'e vectors thus diverges as $x\rightarrow1$. In the small twist regime, the area $A_0$, length $L$ and width $W$ of the moir\'e unit cell read
\begin{align}
A_{0} & =\left|\mathbf{L}_{1}\times\mathbf{L}_{2}\right|=\frac{3\sqrt{3}a^{2}}{2\theta^{2}\left|1-x^{2}\right|},\nonumber\\
L & =\frac{\left|\mathbf{L}_{1}+\mathbf{L}_{2}\right|}{2}=\frac{3a\sqrt{1+\nu x^{2}}}{2\theta\left(1-x^{2}\right)},\nonumber\\
W & =\frac{\left|\mathbf{L}_{1}\times\mathbf{L}_{2}\right|}{\left|\mathbf{L}_{1}+\mathbf{L}_{2}\right|}=\frac{3a}{2\theta\sqrt{1+\nu x^{2}}}.
\end{align}
Although in the critical limit $x\rightarrow1$ both $A_{0}$ and $L$ diverge as $\sim1/\left|1-x\right|$, the width of the unit cell remains finite. 

So far we have considered homobilayer configurations. The generalization to a heterobilayer requires replacing $\mathbf{T}$ by $\mathbf{T}_{h}$, cf. Eq. \eqref{eq:Th}. For the particular case of uniaxial heterostrain in two honeycomb lattices with a lattice mismatch $\sim\epsilon_{b}=\delta$, the critical condition $\mathrm{det}\,\mathbf{T} =0$ yields, to lowest order in $\delta$ and $\theta$, a critical strain magnitude \cite{cosma2014moire}
\begin{equation}
\epsilon_{c}=\frac{\delta\left(1-\nu\right)\pm\sqrt{\left(1+\nu\right)^{2}\delta^{2}+4\nu\theta^{2}}}{2\nu}. 
\label{eq:1D_hetero}
\end{equation}
At low twist angles this reduces to Eq. \eqref{eq:1D_uniaxial} in the homobilayer case ($\delta=0$). As before, this critical strain is independent of the strain angle $\phi$. The above critical strain becomes relevant in moir\'e heterostructures of, e.g., graphene on top of hBN \cite{yankowitz2012emergence, xue2011scanning, decker2011local}, where $\delta\sim0.016$ gives the lattice mismatch between both layers \cite{cea2020band}. This lattice mismatch actually allows Eq. \eqref{eq:1D_hetero} to be satisfied even without a twist, which implies a minimum strain magnitude $\left|\epsilon_{c}\right|\sim\delta$ \cite{cosma2014moire}, see Figure \ref{fig:Strain_Homobilayer_Heterobilayer}(a). 

\begin{figure*}[t]
\centering
    \includegraphics[width=0.9\linewidth]{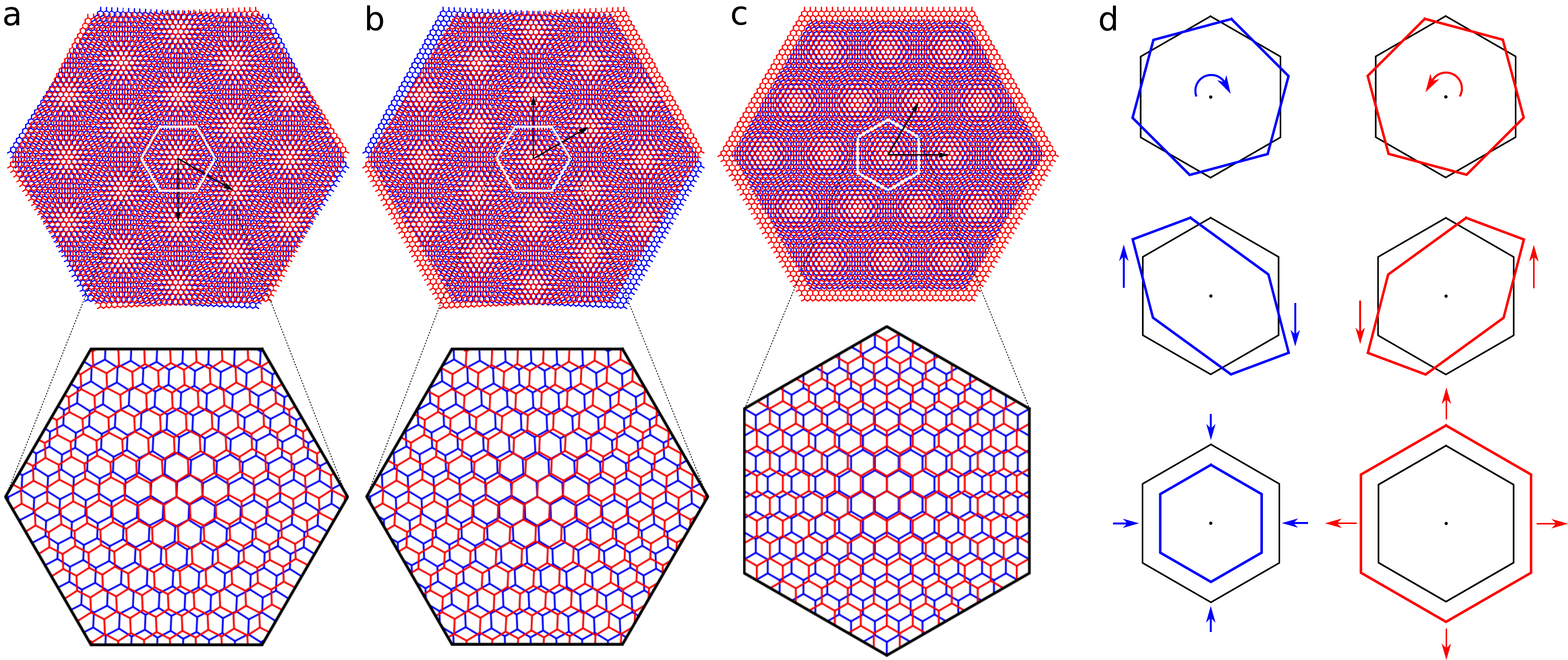}
    \caption{Three ways to generate hexagonal moir\'e patterns in a hexagonal homobilayer configuration: (a) only twist angle, (b) only shear strain and (c) only biaxial strain. The three examples correspond to a twist $\theta=5^{\circ}$, shear strain $\epsilon_{s}=2\sin\left(\theta/2\right)\approx8.7\%$ and biaxial strain $\epsilon_{b}=2\sin\left(\theta/2\right)\approx8.7\%$. By Eq. \eqref{eq:strain_hex}, the three cases have the same moir\'e periodicity. Below each moir\'e pattern it is shown a zoom of the twisted or strained hexagonal layers within the moir\'e unit cell, highlighting the markedly differences in the atomic position of each case, despite their similarities at the moir\'e scale. Panel (d) schematically shows the twist or strain effect in the top and bottom layers (twist, shear and biaxial strain, from top to bottom); the effects are exaggerated for better visualization. Adapted under the terms of the CC BY license from Ref. \cite{escudero2024designing}. Copyright (2024) by the American Physical Society.}
\label{fig:Moire_Hexagonal}
\end{figure*}

\subsubsection{Square patterns}
Another interesting case is the formation of square patterns. These have been predicted theoretically \cite{kogl2023moire, escudero2024designing} (see Figures \ref{fig:Moire_Patterns} and \ref{fig:Moire_Uniaxial_Shear}), and observed experimentally \cite{mendoza2021strain, tilak2022moire, yu2022tunable} (see Section \ref{sec:Strain_Implementation}). In general there is a family of twist and strain configurations that lead to square moir\'e patterns (see, e.g., label II in Figure \ref{fig:Moire_Uniaxial_Shear}). Geometrically, they can be identified by when the moir\'e vectors have equal length and are perpendicular. Within the formalism of Ref. \cite{escudero2024designing} (Section \ref{sec:homobilayers}), the square condition can be stated in term of the $\mathbf{F}$ matrix as [cf. Eq. \eqref{eq:T}]
\begin{align}
\left(\mathbf{F}\mathbf{b}_{1}\right)\cdot\mathbf{b}_{1} & =\left(\mathbf{F}\mathbf{b}_{2}\right)\cdot\mathbf{b}_{2},\nonumber\\
\left(\mathbf{F}\mathbf{b}_{1}\right)\cdot\mathbf{b}_{2} & =0.
\end{align}
Taking the reciprocal vectors given by Eq. \eqref{eq:reciprocal}, the above conditions determine the form of the $\mathbf{F}$ matrix, from which one can then deduce \textit{all} the twist and strain configurations that lead to square patterns. For uniaxial strain in the symmetric configuration ($\theta_{\pm}=\pm\theta/2$ and $\mathcal{E}_{\pm}=\pm\mathcal{E}/2$), the strain magnitude and direction of the square solutions are obtained from Eqs. \eqref{eq:strain_equal} and \eqref{eq:phi_equal} with $\beta=90^{\circ}$. Taking the $r=-1$ root, the needed strain parameters for a given twist angle $\theta$ read 
\begin{align}
\epsilon_{\mathrm{sq}} & =\frac{4}{1-\nu}\frac{\tan\left(\theta/2\right)}{\sqrt{\left(\frac{1-\nu}{1+\nu}\right)^{2}\left(7+4\sqrt{3}\right)-1}},\nonumber\\
\phi_{\mathrm{sq}} & =\frac{\pi}{6}-\frac{1}{2}\arccos\left[\frac{1-\nu}{1+\nu}\left(2+\sqrt{3}\right)\right].
\end{align}
At low twist angle $\tan\left(\theta/2\right)\sim\theta/2$, so $\epsilon_{\mathrm{sq}}\propto\theta$, in agreement with Figure \ref{fig:Moire_Uniaxial_Shear} (as noted in Section \ref{subsec:uniaxial}, this linear dependence at low $\theta$ occurs for any angle $\beta$ between the moir\'e vectors). For $\nu=0.16$, one has $\epsilon_{\mathrm{sq}}\approx0.94\tan\left(\theta/2\right)$ and $\phi_{\mathrm{sq}}\approx-9.4^{\circ}$. At low twist angles ($\theta<5\text{°}$), the needed strain magnitude is thus relatively small; close to the magic angle ($\theta\sim1\text{°}$) one has $\epsilon_{\mathrm{sq}}\sim0.8\%$. A square moir\'e pattern formed by uniaxial strain is shown in Figure \ref{fig:Moire_Patterns}. In Ref. \cite{yu2022tunable} a quasi-square moir\'e pattern is observed in TBG at $\theta=0.38^{\circ}$ with uniaxial heterostrain $\epsilon\sim\left(0.21\pm0.12\right)\%$, in good agreement with the theoretical prediction $\epsilon_{\mathrm{sq}}\left(\theta=0.38^{\circ}\right)\approx0.3\%$ \cite{escudero2024designing}. 

\subsubsection{Hexagonal patterns}\label{sec:triangular_patterns}

Hexagonal moir\'e patterns emerge naturally in the case of only twist without strain (e.g., in twisted bilayer graphene) \cite{andrei2020graphene, andrei2021marvels}. But one can also have hexagonal moir\'e patterns with twist and strain \cite{escudero2024designing}. The simplest case is, of course, with twist and biaxial strain, as this always preserves the hexagonal shape or the layers (Section \ref{sec:Biaxial}). For a strain that deforms the layers, one can still have hexagonal patterns if the strain changes the angle between the moir\'e vector to $\beta=60^{\circ}$, as this leads to the same periodicity as with only a twist (where $\beta=120^{\circ}$). The crucial difference is that the AA stacking regimes are stretched \cite{bi2019designing, escudero2024designing}, so the moir\'e pattern will actually look different (the larger the strain, the more different); see panel Panel (b) of Figure \ref{fig:Moire_Patterns}. In the particular case of uniaxial strain, the hexagonal moir\'e solutions are obtained from Eqs. \eqref{eq:strain_equal} and \eqref{eq:phi_equal} with $\beta=60^{0}$
\begin{align}
\epsilon_{\mathrm{hex}} & =\frac{4\tan\left(\theta/2\right)}{\sqrt{\left(3+\nu\right)\left(1+3\nu\right)}},\nonumber\\
\phi_{\mathrm{hex}} & =\frac{\pi}{6}-\frac{1}{2}\arccos\left(\frac{1}{1+\nu}-\frac{1}{2}\right).
\end{align}
For $\nu=0.16$, one has $\epsilon_{\mathrm{hex}}\approx1.85\tan\left(\theta/2\right)$ --almost twice the needed for the square moir\'e case-- and $\phi_{\mathrm{sq}}\approx-4.38^{\circ}$. Similar solutions can be found for the case of twist and shear strain.

Another configuration that yields hexagonal patterns is with only biaxial or shear strain \cite{cazeaux2023relaxation, escudero2024designing}. Indeed, in both cases the transformation
$\mathbf{F}$ is reduced to $\mathbf{F}=\epsilon^{2}\mathbb{I}$ [cf. Eqs. \eqref{eq:F_shear} and \eqref{eq:F_biaxial}], where $\epsilon$ is the shear or biaxial strain magnitude. As this transformation $\mathbf{F}$ is a spherical tensor, it leads to hexagonal moir\'e patterns. Therefore, pure biaxial or shear strain offer a pathway to engineer hexagonal moir\'e patterns solely by strain. In the case of pure biaxial strain, the moir\'e vectors are simply rescaled by the strain magnitude, $\mathbf{G}_{i}=\epsilon_{b}\mathbf{b}_{i}$, so the moir\'e unit cell has the same orientation as the layers unit cell. For shear strain, the orientation of the hexagonal pattern depends on the shear angle $\varphi$.

By comparing the only strain transformation $\mathbf{F}=\epsilon^{2}\mathbb{I}$ with the transformation $\mathbf{F}=4\sin^{2}\left(\theta/2\right)\mathbb{I}$ with only a twist [Eq. \eqref{eq:F_0}], one readily sees that both have the same periodicity when
\begin{equation}
\epsilon=2\sin\frac{\theta}{2}\sim\theta.\label{eq:strain_hex}
\end{equation}
At low $\theta$, the above strain magnitude is relatively small and within the experimental range. For instance, an equivalence with the TBG magic angle $\theta\sim1.05^{\circ}$ requires $\epsilon\sim1.8\%$. 

Figure \ref{fig:Moire_Hexagonal} shows three hexagonal moir\'e patterns, realized by only a twist, only shear strain and only biaxial strain. All cases correspond to the same moir\'e periodicity. Although at large moir\'e scales the three moir\'e patterns look almost indistinguishable, the actual atomic displacements in each case are markedly different. This turns out to have a significant impact on the electronic properties of the system \cite{escudero2024designing}. In particular, flat bands around a magic twist angle (equivalent or not) \textit{only} emerge in the no-strain configuration. 

\subsection{Moir\'e Brillouin zone}

The variation of the moir\'e vectors due to the strain --specifically, their relative length and angle-- changes the shape of the supercell unit cell. The (first) moir\'e Brillouin zone (mBZ) is, by definition, the particular unit cell in reciprocal space whose borders are closer to the origin than to the others moir\'e vectors. The counterpart in real space is the Winger-Seitz cell (shown in Figure \ref{fig:Moire_Patterns}). 

\begin{figure}[t]
\centering
    \includegraphics[width=\linewidth]{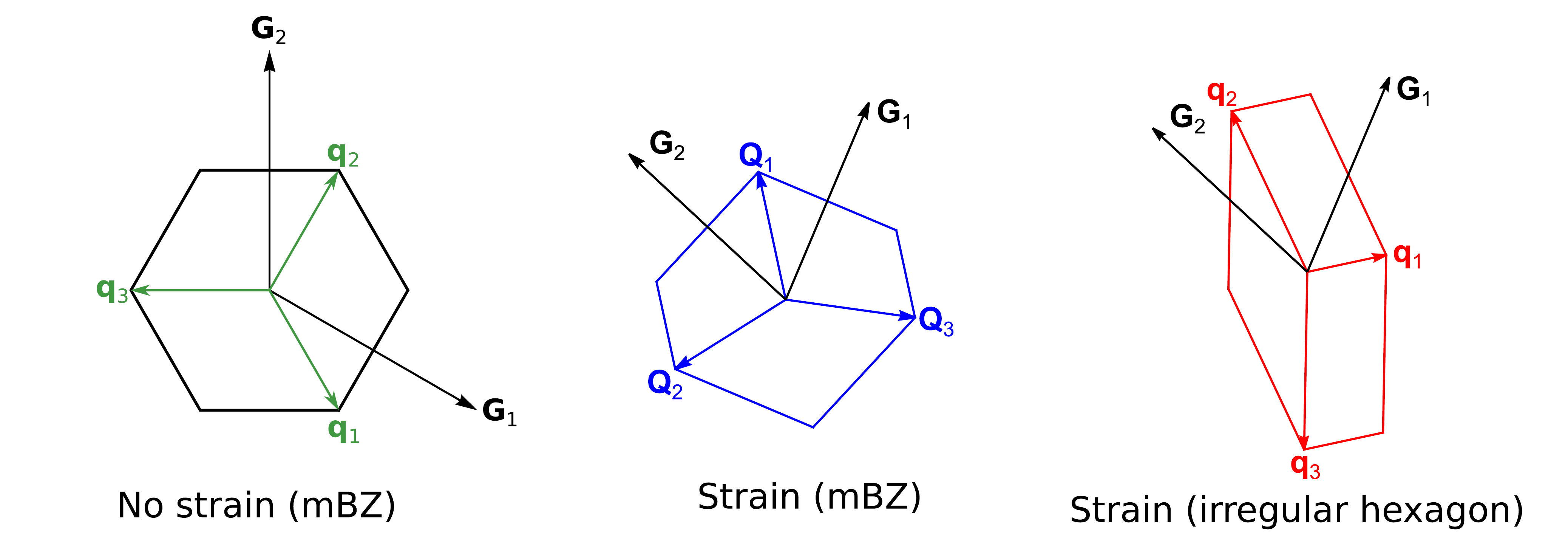}
    \caption{Construction of the moir\'e Brillouin zone (mBZ) with and without strain. The left panel shows the no strain case (only twist) in which the moir\'e pattern is always hexagonal and the mBZ is determined, in terms of the moir\'e vectors $\mathbf{G}_{1}$ and $\mathbf{G}_{2}$, by the three points $\mathbf{q}_{i}$ given by Eq. \eqref{eq:qi_nostrain}. The other two panels show the case with twist and strain, whereby the moir\'e pattern is no longer hexagonal and the correct mBZ is determined, for equal-length moir\'e vectors, by the points $\mathbf{Q}_{i}$ given by Eq. \eqref{eq:qi_nostrain} (middle panel). With strain, the points $\mathbf{q}_{i}$ rather define a stretched irregular hexagon (right panel) which is not the true (first) mBZ.}
\label{fig:mBZ_Deformation}
\end{figure}

\begin{figure}[t]
\centering
    \includegraphics[width=\linewidth]{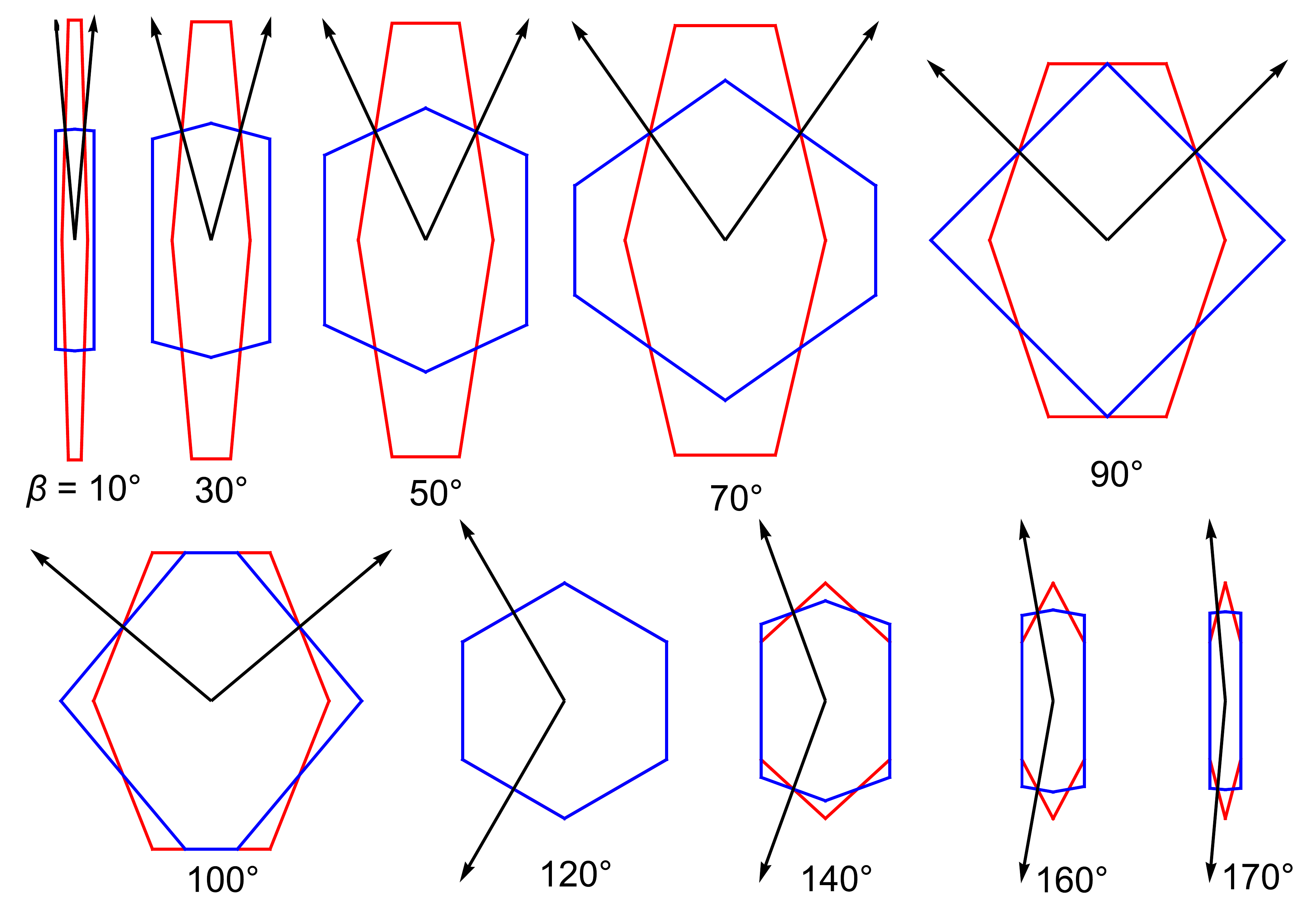}
    \caption{Evolution of the moir\'e Brillouin zone (mBZ) for different angles $\beta$ between equal-length moir\'e vectors $\mathbf{G}_{i}$ (black arrows). For each angle $\beta$, the correct mBZ defined by the points $\mathbf{Q}_{i}$ of Eq. \eqref{eq:qi_strain} is shown in blue, while the stretched irregular hexagon defined by the points $\mathbf{q}_{i}$ of Eq. \eqref{eq:qi_nostrain} is shown in red. Only in the no strain case of $\beta=120^{\circ}$ do both cells coincide. For nonzero strain, only the correct mBZ reflects the periodicity of the moir\'e pattern. Adapted under the terms of the CC BY license from Ref. \cite{escudero2024designing}. Copyright (2024) by the American Physical Society.}
\label{fig:mBZ_Evolution}
\end{figure}

In the the case of only a twist with no strain, the moir\'e vector have always a $120^{\circ}$ angle between them, and the moir\'e pattern is triangular. The corresponding mBZ is thus a hexagon, and it is straightforward to show that its six borders can be determined from the moir\'e vectors by the points \cite{moon2013optical, koshino2018maximally}
\begin{align}
\mathbf{q}_{1} & =-\frac{2\mathbf{G}_{1}+\mathbf{G}_{2}}{3},\nonumber \\
\mathbf{q}_{2} & =\mathbf{q}_{1}+\mathbf{G}_{1},\nonumber \\
\mathbf{q}_{3} & =\mathbf{q}_{1}+\mathbf{G}_{1}+\mathbf{G}_{2}.
\label{eq:qi_nostrain}
\end{align}
and their negatives (see Figure \ref{fig:mBZ_Deformation}). These points determine the correct mBZ only in the specific case of equal length moir\'e vectors with an angle $120^{\circ}$ between them. 

With strain, the moir\'e vectors do not longer keep, in general, the same relative length and direction, so the cell determined by the points given by Eq. \eqref{eq:qi_nostrain} is no longer the true mBZ of the heterostructure. Although the points $\mathbf{q}_{i}$ would still define a unit cell in reciprocal space (namely, a stretched hexagon; see Figure \ref{fig:mBZ_Deformation}), it would not reflect the change in the moir\'e geometry due to the interplay with twist and strain. 

In general, the correct mBZ must be constructed for the specific twist and strain (in practice, numerically). The construction is greatly simplified in the particular case of twist and strain that give equal length moir\'e vectors (see Section \ref{subsec:uniaxial}). In that case, the correct points that determine the mBZ are given by \cite{escudero2024designing}
\begin{align}
\mathbf{Q}_{1} & =-\frac{\left(1+2\chi\right)\mathbf{G}_{1}-\lambda\mathbf{G}_{2}}{2\left(1+\chi\right)},\nonumber \\
\mathbf{Q}_{2} & =\mathbf{Q}_{1}+\mathbf{G}_{1},\nonumber \\
\mathbf{Q}_{2} & =\mathbf{Q}_{1}+\mathbf{G}_{1}-\lambda\mathbf{G}_{2}.
\label{eq:qi_strain}
\end{align}
and their negatives. Here
\begin{equation}
\chi=\frac{\left|\mathbf{G}_{1}\cdot\mathbf{G}_{2}\right|}{\left|\mathbf{G}_{1}\cdot\mathbf{G}_{1}\right|},\qquad\lambda=\mathrm{sign}\left(\mathbf{G}_{1}\cdot\mathbf{G}_{2}\right)+\delta_{\mathbf{0},\mathbf{G}_{1}\cdot\mathbf{G}_{2}}.
\end{equation}
Eq. \eqref{eq:qi_strain} is a generalization of Eq. \eqref{eq:qi_nostrain} for any angle between equal-length moir\'e vectors (to which it reduces when $120^{\circ}$). In the special case of $\beta=90^{\circ}$ (square moir\'e pattern), the four points reduce to four because $\mathbf{Q}_{1}=-\mathbf{Q}_{3}$. Note that Eq. \eqref{eq:qi_strain} determines the borders of any two-equal length lattice vectors with arbitrary angle, so they also determine the Wigner-Seitz cell in real space (using the equal-length moir\'e vectors $\mathbf{G}_{i}^{R}$). Figure  \ref{fig:mBZ_Evolution} shows the evolution of the mBZ for equal length moir\'e vectors with different angle $\beta$. 

The mBZ depicts the symmetry of the twisted and strained moir\'e vectors. In conventional band theory, the borders of the BZ represent special, high-symmetries points because they are located at the middle between two (or more) Bragg planes, at which a periodic potential opens a gap and there is a constructive interference condition (Bragg reflection). In nonstrained moir\'e heterostructures, the system preserves the symmetries of the underlying hexagonal lattices, and the borders of the mBZ still represent high-symmetry points. However, under strain the system loses most of its symmetries and the borders of the mBZ do no longer represent, in general, special high-symmetries points. In particular, under strain the positions of the two Dirac points are no longer located at the vertices of the mBZ \cite{naumis2017electronic, bi2019designing, escudero2024designing}; the Dirac points are rather located at unfixed points within the mBZ (their exact position depends on the specific twist and strain parameters). Therefore, unlike conventional band theory in symmetric crystal structures, the borders of the twisted and strained mBZ do not necessarily reflect special points of the moir\'e band structure (e.g., where the Dirac points are located). The correct shape of the mBZ can still be important in interpreting, and correctly explaining, experimental transport measurements \cite{wang2023unusual}.

\subsection{Lattice relaxation}\label{sec:lattice_relaxation}

The geometrical description in the previous sections assumes that the lattices deform following the externally applied twist and strain. Real systems, however, tend to naturally relax in response to external perturbations, in order to minimize the energy. Relaxation effects in moir\'e heterostructures have been extensively studied in the literature \cite{van2015relaxation, nam2017lattice, zhang2018structural, carr2018relaxation, guinea2019continuum, lucignano2019crucial, cantele2020structural, leconte2022relaxation, ezzi2024analytical, kang2025analytical}. 
In the case of twisted bilayer graphene, the relaxation is particularly significant at very small (marginal) twist angles \cite{van2015relaxation}, whereby the energy-costly AA stacking regimes shrink and corrugate out of plane, leading to the formation of triangular AB/BA stacking separated by domain walls (DW) \cite{zhang2018structural}. As a result, the relaxed moir\'e pattern of slightly twisted bilayer graphene can look markedly different than those expected for a rigid configuration. The relaxation acts, by itself, as a source of intrinsic inhomogeneous strain in the system \cite{kazmierczak2021strain, kang2023pseudomagnetic, ceferino2024pseudomagnetic}.

Similar relaxation effects are observed under external strain, although to date only few works have studied it. In the particular case of triangular moir\'e patterns induced by solely biaxial or shear strain (Section \ref{sec:triangular_patterns}), Cazeaux \textit{et al.} reported also an expansion and contraction of AB/BA and AA stacking regimes, respectively, as the strain magnitude decreases \cite{cazeaux2023relaxation}. A similar behavior was also recently reported by Kundu \textit{et al.} in biaxially strained $\mathrm{MoS_{2}}$ bilayer \cite{kundu2025atomic}. There, the relaxed in-plane strain distribution is also modified by reducing and increasing, respectively, the strain magnitudes at the AB/BA and AA stacking regimes, which as in twisted systems, are separated by domain walls. Since these relaxation effects arise due to local minimization of the stacking and elastic energies, a similar behavior of DW formation is generally expected for non-triangular strained patterns. For instance, the quasi-unidimensional channels (Section \ref{sec:1D_channels}) also undergo a shrink and expansion of the AA and AB/BA stackings \cite{cazeaux2023relaxation}. Another interesting relaxation effect is the formation of giant atomic swirls under pure biaxial heterostrain \cite{mesple2023giant}, whereby the soliton DW swirl around the AA stacking regimes \cite{lalmi2014flower}. 

In general, the relaxed moir\'e configurations cannot be accounted by homogeneous twist and strain profiles, as considered in previous sections. Rather, they are characterized by a position dependent displacement vector at each atomic site, which minimizes the elastic and interlayer adhesion energies. The displacement vectors are typically obtained numerically \cite{nam2017lattice, carr2018relaxation, cazeaux2020energy}, or analytically \cite{vafek2023continuum, ezzi2024analytical, kang2025analytical}. The geometrical description of relaxed moir\'e patterns arising from inhomogeneous strain is beyond the scope of this review.

\subsection{Experimental determination of the strain}
It is often the case that the strain in moiré heterostructures is not induced or controlled (Sec. \ref{sec:Strain_Implementation}), but rather arise naturally during the sample fabrication. In such situations, the strain configuration in the system is not known beforehand, so it can only be inferred by analyzing the geometrical and electronic properties of the moiré pattern. In this section, we briefly describe several methods to identify the strain and twist from the scanning tunneling microscopy (STM) topography of moiré structures \cite{hermann2012periodic,artaud2016universal,huder2018electronic,kogl2023moire,kerelsky2019maximized,yu2024twist,carrasco2025}. 

A general formalism to identify the strain and twist from the atomically-resolved microscopy of any hexagonal moiré materials was first proposed by Artaud \textit{et al.} \cite{artaud2016universal}. In this method, a set of eight integers is identified by a Fourier analysis of STM images, from which one can then use the extended Wood's notation to determine the geometry of the moiré superlattice. By using this method, Huder \textit{et al.} \cite{huder2018electronic} found that a small uniaxial heterostrain can suppress Dirac cones and induce flat bands in TBG. Kerelsky \textit{et al.} \cite{kerelsky2019maximized} proposed another method to deal with STM images without atomic resolution, whereby by assuming that the samples suffer a uniaxial heterstrain one could extract the twist, strain magnitude and strain direction with the information of the moiré wavelengths only. Recently, Yu \textit{et al.} \cite{yu2024twist} and Carrasco \textit{et al.} \cite{carrasco2025} develop a simple minimization procedure that determines the twist and strain without presuming whether the strain is uniaxial or shear. By comparing the measured moir\'e wavelengths with the theoretical expressions for both strain types, the method identifies the twist-strain configuration that best matches the experimental results, providing a systematic way to determine the dominant strain component and the corresponding parameters directly from STM data. 

All the above methods aim to infer the twist and strain configuration in the system through only the in-plane position of the atoms. The strain effect on the interlayer distance of the moiré pattern can be directly estimated from the STM topographic height profile \cite{kerelsky2019maximized}.  

\begin{figure*}[t]
\centering
    \includegraphics[width=0.8\linewidth]{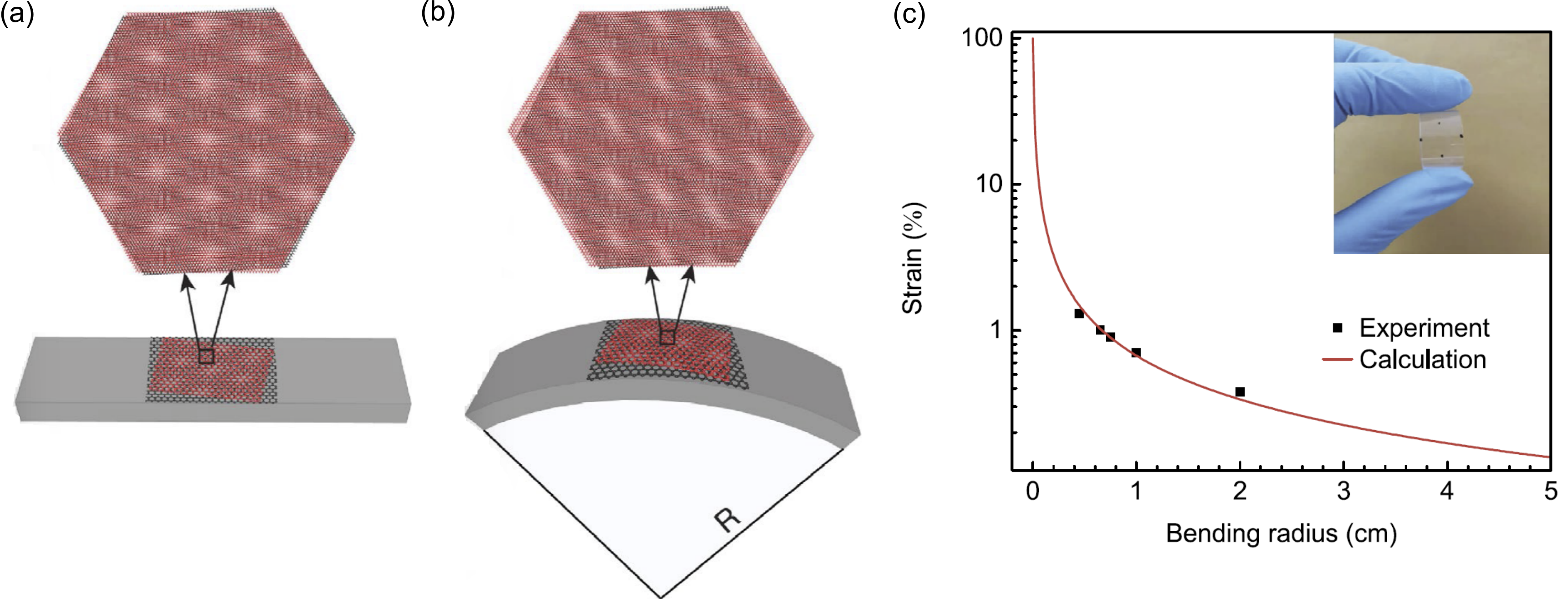}
    \caption{The substrate bending technique. Schematic illustration of generating heterostrain in TBG on a flexible substrate (a) before and (b) after bending. R is the bending radius. (c) Calculated (red line) strain on the flexible substrate surface and experimental (black squares) uniaxial strain in bottom graphene. The thickness of the flexible substrate is 135 $\mathbf{\mu}$m. Inset shows the flexibility of the substrate with TBG device on top. Adapted under the terms of the CC BY license from Ref. \cite{gao2021heterostrain}. Copyright (2021) The author(s).}
\label{figs4_bending}
\end{figure*}

\section{Experimental implementation of strain in moir\'e materials}\label{sec:Strain_Implementation}

Due to the nature of nanoscale dimensions and stability, strain is a powerful tuning knob to modify the geometrical and electronic properties of 2D systems \cite{lee2013high,li2015optoelectronic,zhang2018strain,so2021polarization,darlington2020imaging,levy2010strain,bunch2008impermeable,lloyd2016band,thai2021mos2,conley2013bandgap,datye2022strain,castellanos2013local,deng2016wrinkled,lee2020switchable}. In particular, as we discuss in Sec. \ref{sec:twist_strain}, due to the presence of a moir\'e that acts as a magnifying glass, even very small strains can significantly modify the geometrical and electronic properties of the moir\'e patterns. In practice, small strain tend to be induced inevitably without control during the sample growth or fabrication, which we refer to as unintentional strain. Intentional strain may be also induced externally and designed carefully by well-established experimental techniques. In this section, we introduce three efficient experimental techniques developed to engineer strain in moir\'e materials. Then we present various strain-induced moir\'e pattern geometries generated in experiments. 
\begin{table}[t]
\small
\setlength{\tabcolsep}{3pt}\renewcommand{\arraystretch}{1}

\begin{tabular*}{0.49\textwidth}{@{\extracolsep{\fill}}%
  >{\raggedright\arraybackslash}p{0.15\textwidth}%
  >{\raggedright\arraybackslash}p{0.12\textwidth}%
  >{\raggedright\arraybackslash}p{0.21\textwidth}@{}}
\hline
\textbf{Technique} & \textbf{Strain range} &
\textbf{Moiré materials} \\
[2pt]
\hline
Substrate out-of-plane bending 
& up to $\sim2.5\%$
& TBG \cite{gao2021heterostrain,liu2024continuously}, Graphene/hBN \cite{kapfer2023programming,wang2019situ},
Twisted WSe$_2$ \cite{ou2025continuous}. \\
[3pt]
Process-induced strain 
& up to $\sim 3\%$
&  TBG \cite{pena2023moire}, 
Twisted MoS$_2$/WSe$_2$ \cite{zhang2024patternable}, Bilayer MoSe$_2$ \cite{edelberg2020tunable}. \\
[3pt]
Sliding-based strain 
& up to $\sim 0.81\%$
& Graphene/hBN \cite{sequeira2024manipulating},
TBG \cite{carrasco2025}. \\
\hline
\end{tabular*}
\caption{Summary of strain implementation techniques, including the strain range and applicable moiré materials, discussed in Sec. \ref{sec:strain_techniques}.}
\label{table1}
\end{table}

\subsection{Strain techniques}\label{sec:strain_techniques}

A large amount of experimental techniques have been introduced to generate strain in 2D materials. For example, strain can be generated by using different microactuators \cite{perez2014controlled,ziss2017comparison}, by applying hydrostatic pressure \cite{antoniazzi2024pressure}, by placing 2D materials on periodic structures constructed by nanopillars \cite{li2015optoelectronic,jiang2017visualizing}, by stacking on piezoelectric stacks \cite{liu2024continuously}, or by bending a flexible substrate \cite{wang2019situ}. Here, we will mainly focus on several common approaches that are well-implemented in moir\'e materials. A brief summary of the key strain implementation techniques for engineering moiré materials is given in Table \ref{table1}. Note that the strain techniques developed for moir\'e materials may not be compatible for moir\'e material-based devices. The strain approaches for 2D material-based devices are out of the scope of this review.

\subsubsection{Substrate out-of-plane bending}
\label{bending}

\begin{figure*}[t]
\centering
    \includegraphics[width=0.8\linewidth]{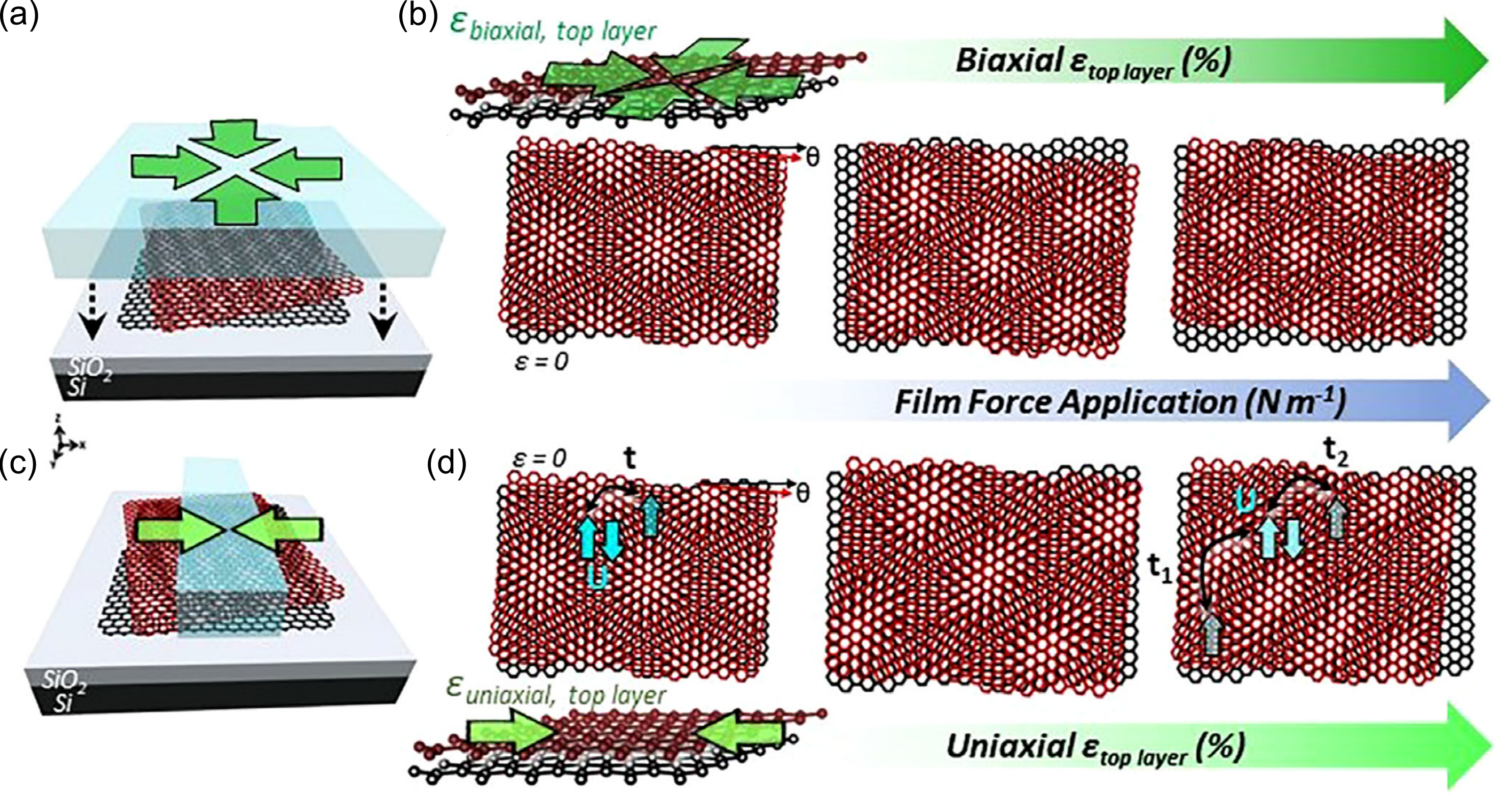}
    \caption{Process-induced strain technique. (a) Schematic illustration of generating biaxial heterostrain by fully encapsulating a TBG with a highly stressed thin film. (b) Moir\'e patterns variation with biaxial heterostrain strength, where $\theta$ is the twist angle in the TBG. (c) Schematic illustration of generating uniaxial heterostrain by partially encapsulating TBG with a highly stressed thin film patterned into a stripe. (d) Moir\'e patterns variation with uniaxial heterostrain strength, and strain tunable Hubbard model parameters. The Hubbard model parameters $U$ and $t$ are the Coulomb repulsion and hopping terms, respectively. In the process-induced strain approach, the strain magnitude is directly proportional to the applied thin force on the stressor layer. Reprinted from Ref. \cite{pena2023moire}, with the permission of AIP Publishing.}
\label{figs4_stressor}
\end{figure*}

\begin{figure*}[t]
\centering
    \includegraphics[width=0.8\linewidth]{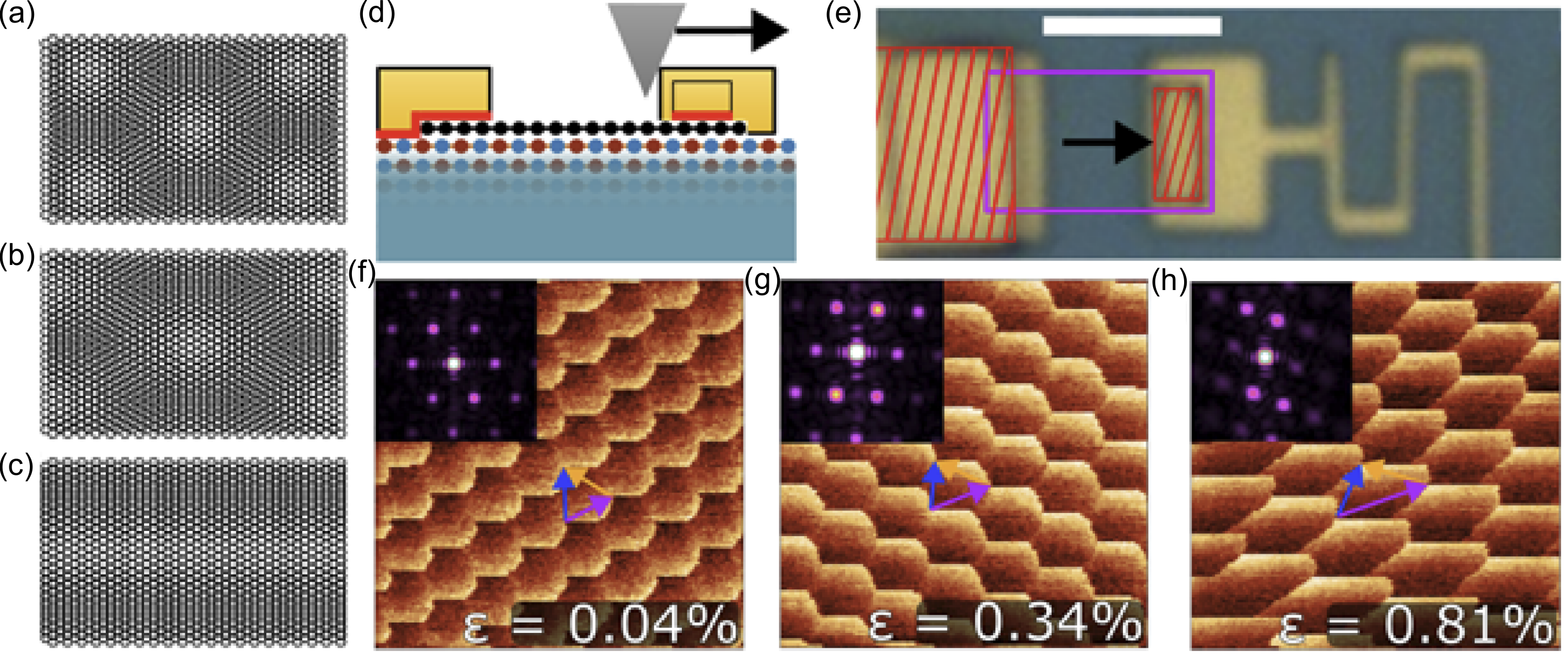}
    \caption{Sliding-based strain technique. (a)-(c) Moir\'e patterns generated by hexagonal lattices with a lattice mismatch $\delta=10\%$ and (a) no strain, (b) $\varepsilon=2$\% uniaxial heterostrain, (c) $\varepsilon=10$\% uniaxial heterostrain. (d) Slide view schematic of a strained open-face graphene-hBN device. The graphene is stretched by pushing the right electrode with an AFM tip (gray triangle). Red lines show regions where the metal is strongly adhered through the O$_2$--plasma treatment before deposition. (e) Optical image of a device. Red dashed regions are the O$_2$--plasma treated electrode. Purple box labels the graphene channel. Scale bar is 4 $\mathbf{\mu}$m. (f)-(h) 100 nm $\times$ 100 nm CAFM images of graphene-hBN moir\'e patterns. The $x$-axis of the CAFM is along the graphene channel, parallel to the stretching direction. Insets show FFT of the images. Fitting the wavevectors results in the strain values $\varepsilon$ with strain angles $\phi$ relative to the $+x$-axis, and twist angle $\theta$ between graphene and hBN as (0.04\%, +80 $\pm$ 20$^\circ$, -0.12$^\circ$), (0.34\%, +1 $\pm$ 5$^\circ$, 0.02$^\circ$) and (0.04\%, +17 $\pm$ 4$^\circ$, 0.02$^\circ$) for (f), (g) and (h), respectively. Reprinted with permission from Ref. \cite{sequeira2024manipulating}. Copyright 2024 American Chemical Society.}
\label{figs4_sliding}
\end{figure*}

One strain technique for moir\'e materials is based on in-plane or out-of-plane stretching or bending the underlying substrate or active 2D materials  \cite{wang2019situ,gao2021heterostrain,kapfer2023programming,ou2025continuous}. In this technique, the active 2D layer can be suspended or directly stacked on the substrate. We refer to this method as substrate bending technique. A similar method to generate strain, through in-plane bending of one active 2D layer by the tip of an atomic force microscope (AFM), will be detailed in the following subsection \cite{kapfer2023programming}. 

Figure \ref{figs4_bending} presents a generalized approach of dynamically tuning heterostrain in TBG via the substrate out-of-plane bending \cite{gao2021heterostrain}. The setup consists of TBG assembled on a flexible polymethyl methacrylate (PMMA)-coated polyethylene terephthalate (PET) substrate. The twist angle between the two graphene layers is large (about $\theta \sim 13.2^{\circ}$ in the experiment) in order to minimize the interlayer friction force in TBG. As a result, the bottom graphene layer is separated from both the top graphene and the flexible substrate, and the strain generated from the substrate is only transferred to the bottom graphene. Upon bending the PET substrate, because the bottom graphene layer is strongly adhered to the substrate, it becomes uniaxially stretched. The top graphene experiences a negligible amount of deformation because it is bonded by the weak van der Waals force. Consequently, a uniaxial heterostrain is generated in the TBG moir\'e pattern. The strain strength can be dynamically and \textit{in situ} modulated by changing the bending radius $R$, following the relation $\epsilon \approx t/2R$, where $t$ is the substrate thickness (Fig. \ref{figs4_bending}(b)).  The maximum strain strength achieved by out-of-plane bending in TBG with $\theta \sim 13.2^{\circ}$ was $\sim 1.3\%$ (Fig. \ref{figs4_bending}(c)). The formation of heterostrain can be confirmed by the Raman spectroscopy, which presents triple G peaks with one degenerate mode G peak from unstrained graphene, and two split G$^+$ and G$^-$ peaks from strained graphene.    

\subsubsection{Process-induced strain}
\label{process}

Another common approach is a process-induced strain technique, which is based on the deposition of stressed thin films onto the active 2D materials  \cite{pena2021strain,pena2023moire,zhang2024enhancing,zhang2024patternable,liu2025programmable}. In Ref. \cite{pena2023moire}, Peña \textit{et al.} utilized this method to generate designable layer-by-layer strains, on the device-level, with full control over the type of strain (uniaxial or biaxial, tensile or compressive) in moir\'e systems. By choosing optically transparent stressors, the intentionally induced strain could be characterized through Raman spectroscopy. 

The technique is schematically shown in Figure \ref{figs4_stressor}(a) and (c). It consists of a thermally evaporated stressor of CrO$_x$ (10 nm) / MgF$_2$ (X nm) fully or partially encapsulated onto the TBG moir\'e pattern, respectively. Due to the weak interlayer coupling in TBG with relatively large twist angle, the compressive strain induced by the tensile film is only transferred into the top graphene layer. The tensile film force is adjusted by changing the thickness of the MgF$_2$ layer alone (40 nm to 125 nm). The strain magnitude is directly proportional to the film force [film stress (GPa) $\times$ film thickness (nm)]. Figures \ref{figs4_stressor}(b) and (d) show that the induced biaxial or uniaxial heterostrain provides a pathway to modify the geometry and symmetry of the moir\'e patterns. Note that in the biaxial heterostrain case the moir\'e pattern effectively retains the C$_3$ rotational symmetry. For uniaxial heterostrain there is rather a significant change in the moir\'e geometry, resulting in nonhexagonal patterns as described in Section \ref{subsec:uniaxial}. The strain and the moir\'e pattern interference can be examined by Raman spectroscopy through in-plane and moir\'e-activated phonon mode shifts. The process-induced strain approach is a powerful technique that allows for many tunable parameters, such as the strain magnitude, tension or compression, uniaxial or biaxial, strain direction, and strain profiles.  

\begin{figure*}[t]
\centering 
    \includegraphics[width=\linewidth]{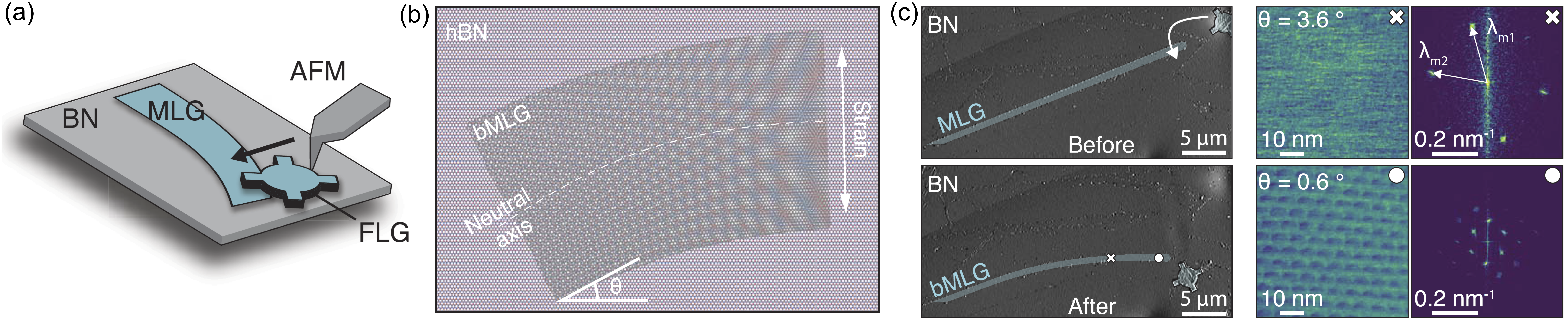}
    \caption{Continuous tuning of twist angle and strain in a graphene--hBN heterostructure. (a) Schematic of the continuous tuning approach. The twist angle and the strain are manipulated by bending a 2D material ribbon through a nanomanipulator and the AFM tip. (b) Sketch of a distorted moir\'e pattern, magnifying the effect of strain in the low-angle limit. (c) (Left panel) AFM images of a bMLG on hBN before (top) and after (bottom) bending, (middle panel) PFM scans taken at the marked spots (white cross and dot) indicated in the bMLG  and (right panel) their corresponding FFTs. From Ref. \cite{kapfer2023programming}. Reprinted with permission from AAAS.}
\label{figs4_bending_out}
\end{figure*}

\subsubsection{Sliding-based strain}
\label{sliding}

In Ref. \cite{sequeira2024manipulating}, Sequeira \textit{et al.} presented a sliding-based technique to manipulate the uniaxial heterostrain in graphene-hexagonal boron nitride (hBN) and study the strain effect on moir\'e geometry. To this end, they fabricated an open-faced graphene device on a hBN substrate, with a fixed source electrode and a movable drain electrode. In the electrode region, a metal handle is strongly adhered to the graphene layer (red line in Fig. \ref{figs4_sliding}(d)). Consequently, the drain electrode can be moved laterally by pushing it with an AFM tip, inducing heterostrain in the graphene layer. To selectively adhere the electrode to only the graphene layer, a light O$_2$-plasma treatment is applied before depositing the Cr/Au handles. The electrodes are used to both selectively strain and measure transport (Fig. \ref{figs4_sliding}(e)). Importantly, the movable electrode has to have sufficient sliding friction to retain the strain in graphene when the AFM tip is retracted. As a result, an open-face graphene sample can be progressively strained, independently of the hBN substrate. The strain effect was measured by performing the conductive AFM (CAFM) measurement on the open graphene channels (Figs. \ref{figs4_sliding}(f)-(h)). As seen, the initial graphene aligned with hBN, forming a moir\'e pattern of length ~ 14 nm, evolves into elongated moir\'e patterns by stretching the graphene with the AFM tip.

Recently, a similar sliding-based strain method was developed by Carrasco \textit{et al.} \cite{carrasco2025}. During moir\'e sample growth, due to the different thermal expansion between graphene and the underlying substrate, graphene layers experience different sources of strain, resulting in graphene wrinkles in the graphene moir\'e sample. By mechanically manipulating the graphene wrinkles with a STM tip, they actively tune local strain in TBG around the angle $\sim 1.13^{\circ}$. The geometry of the moir\'e pattern can be dynamically and reversibly switched between trigonal and square. The square pattern arises from a combination of twist and shear strain that minimizes the elastic energy. 

\begin{figure*}[t]
\centering
    \includegraphics[width=0.8\linewidth]{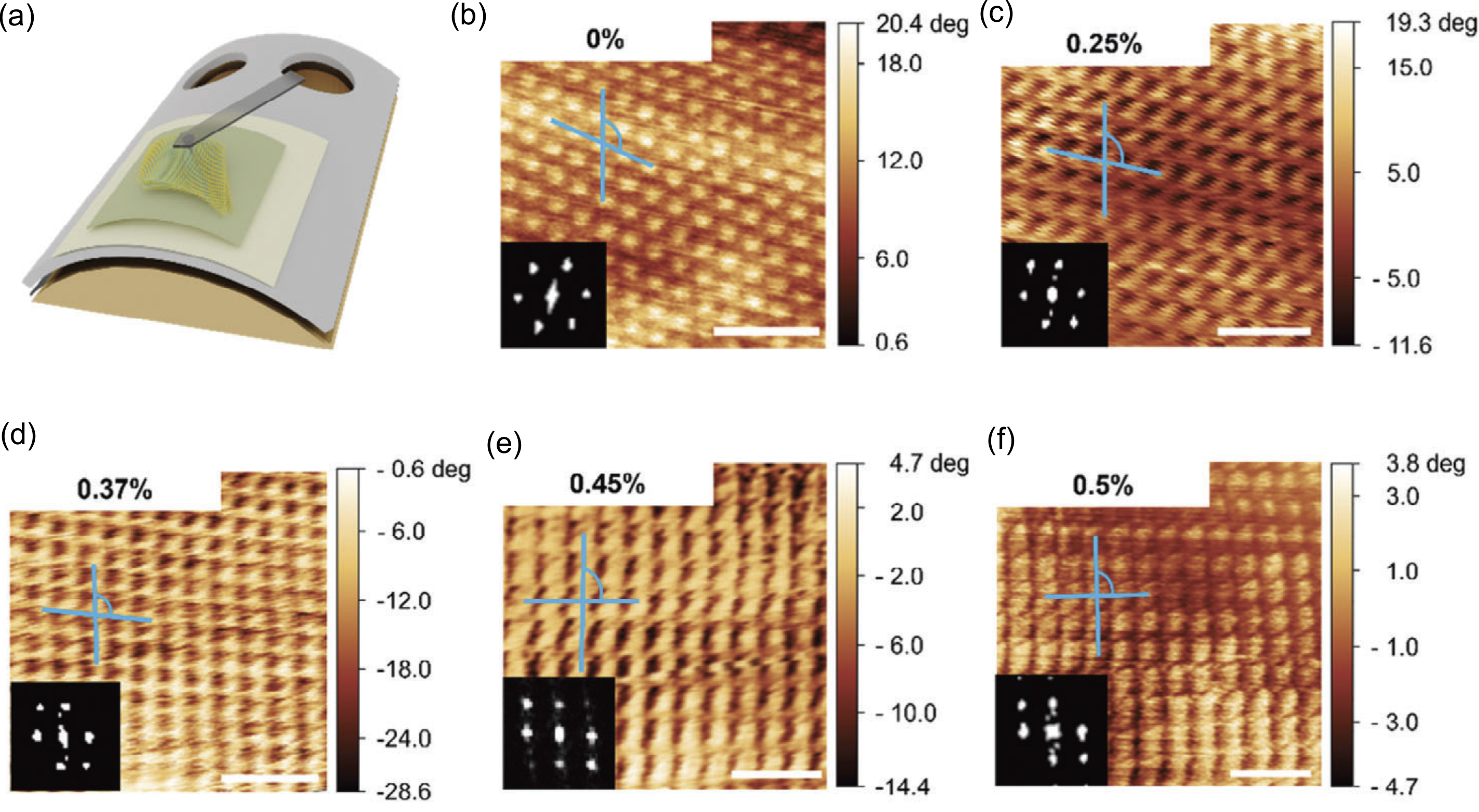}
    \caption{Strain-induced triangular-rectangular moir\'e pattern evolution in twisted bilayer WSe$_2$ with twist angle $\theta\approx1.35^\circ$. (a) Schematic illustration of a piezoresponse force microscopy (PMF) technique to directly visualize moir\'e patterns on bent substrates. The substrate was mounted on molds with different curvatures to generate different strains. (b)-(f) PFM images of twisted bilayer WSe$_2$ with substrate mounted on molds with (b) flat surface, strain magnitude $\varepsilon=0\%$, (c) with radius $R=25$ mm, $\varepsilon=0.25\%$, (d) with radius $R=16.7$ mm, $\varepsilon=0.38\%$, (e) with radius $R=13.9$ mm, $\varepsilon=0.45\%$, and (f) with radius $R=12.5$ mm, $\varepsilon=0.5\%$. Inserts are FFT images. The blue crosses indicate the moir\'e angle $\alpha$. The scale bars are 50 nm. Adapted under the terms of the CC 4.0 license from Ref. \cite{ou2025continuous}. Copyright 2025 The author(s).}
\label{figs4_rectangular}
\end{figure*}

\subsection{Strained moir\'e patterns}

As discussed in previous sections, the strain is a powerful tuning knob for manipulating the size and symmetry of the moir\'e patterns, much beyond what is possible via only a twist angle. In this section, we provide experimentally examples of nonhexagonal moir\'e geometries arising due to combinations of twist and strain. Here, the strain could be intentionally generated through well-developed strain techniques (Section \ref{sec:strain_techniques}), unintentionally induced during the sample fabrication, or appear due to the lattice relaxation. We will first illustrate a continuous approach to fine tune the twist and strain \cite{kapfer2023programming}, and then show particular observations of rectangular patterns \cite{ou2025continuous}, quasi-one-dimensional patterns \cite{alden2013strain,bai2020excitons,mendoza2021strain,shabani2021deep}, and giant atomic swirl deformations \cite{mesple2023giant}.

\subsubsection{Continuous tuning moir\'e patterns}

In the moir\'e systems, the geometric and electronic properties are often highly dependent on the moir\'e periodicity and the interlayer atomic registry. For example, the superconductivity in TBG is observed around a twist angle 1.1$^\circ$, but has a tolerance of only 0.1$^{\circ}$ \cite{cao2021nematicity}. Therefore, a further optimization of the moir\'e electronic properties requires a technique to precisely and continuously tune the moir\'e pattern with twist angle and strain \cite{kapfer2023programming,pantaleon2024designing}. 

Kapfer \textit{et al.} develop a continuous approach to fine tune the twist angle and strain in a graphene--hBN heterostructure, based on the in-plane bending of monolayer ribbons through an AFM tip \cite{kapfer2023programming}. Figure \ref{figs4_bending_out} illustrates the continuous tuning approach: a monolayer graphene (MLG) ribbon is bent by a few-layer graphite (FLG) manipulator shaped into a gear-like geometry and an AFM tip. The strain induced by bending the ribbon varies both in the longitudinal (x) and transverse (y) directions. The strain map in the bent ribbon follows some general tendencies: (\romannum{1})\; the strain changes linearly in both directions and has the maximum value at the end of the ribbon; (\romannum{2})\; in the transverse direction, the ribbon has compressive and tensile strain on the inside and outside radius, respectively, which are separated by an unstrained neutral axis; (\romannum{3})\; the maximum strain in the transverse and longitudinal directions are linearly dependent on the width $W$ and length $L$ of the ribbon, respectively. Therefore, this continuous tuning approach allows one to manipulate independently the twist angle and strain in 2D materials. Importantly, the bending process is robust and reversible.


\subsubsection{Rectangular moir\'e patterns}

Figure \ref{figs4_rectangular} shows a strain-induced moir\'e pattern manipulation in twisted bilayer WSe$_2$ \cite{ou2025continuous}. In this setup, a continuous and homogeneous uniaxial heterostrain is generated by bending a flexible substrate with different radius $R$ (see Sec. \ref{bending}). The deformed moir\'e patterns were visualized by a piezoresponse force microscopy (PFM). As discussed in the previous subsection, in the bending substrate technique the strain is only transferred from the substrate to the bottom layer WSe$_2$, generating a heterostrain in the moir\'e pattern. In the experiment, the substrate was mounted on several molds with different curvatures, which controlled the strain magnitude from 0\% (flat) to 0.5\% (radius curvature of 12.5 mm). As seen in Figures \ref{figs4_rectangular}(b)-(f), this strain variation results in a continuous symmetry modulation of the moir\'e patterns, from triangular ($\beta_R=120^\circ$) to rectangular ($\beta_R=90^\circ$).

\begin{figure}[t]
\centering 
    \includegraphics[width=0.85\linewidth]{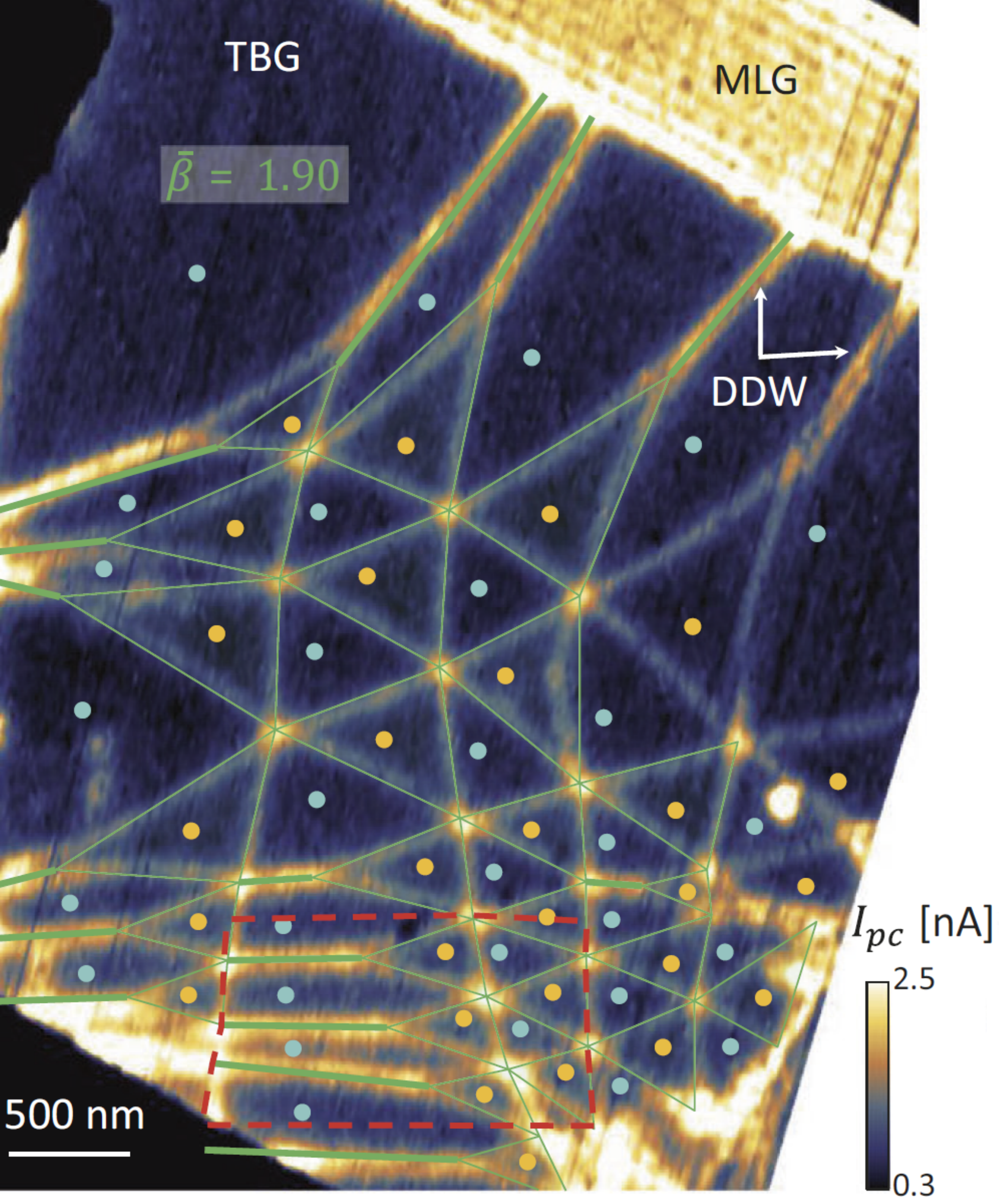}
    \caption{Strain induced 1D pattern in TBG. Non-local nano-photocurrent map of a TBG sample with $\theta <0.1^\circ$. In the photocurrent map, the bright spots highlight the AA sites. The AB and BA regions are indicated by orange and cyan dots, respectively. In the strained region, double domain walls (DDWs) are formed, separating regions of identical stacking configurations (AB or BA stackings). The green network overlaid on the photocurrent data corresponds to the prediction with one tuning parameter $\bar{\beta}$, indicating the ratio of DDW and single domain wall (SDW) line energies. Adapted under the terms of the CC BY license from Ref. \cite{halbertal2021moire}. Copyright (2021) The author(s).}
\label{figs4_1D}
\end{figure}

\subsubsection{Quasi-1D pattern}

As discussed in Section \ref{sec:1D_channels}, in strained moir\'e materials there can be a critical strain configuration at which the moir\'e collapse into a quasi-1D pattern. The critical condition can be, in principle, engineered by extenally induced-strain (Section \ref{sec:strain_techniques}). But quasi-1D pattern can also emerge most naturally by unintensional strains, either due to lattice relaxation or sample fabrication. For instance, Figure \ref{figs4_1D} shows clear rich features of quasi-1D moir\'e strips of relaxed DWs, for example, the shapes of domains, single domain walls (SDWs), and formation of double domain walls (DDWs) \cite{halbertal2021moire}. A dimensionless parameter as the ratio of DDW and SDW line energies was defined to adress a competition between SDWs and DDWs. For $\bar{\beta}<2$, two SDWs attracted each other favoring the emergence of DDW segments. The SDWs and DDWs are also observed in some other moir\'e materials \cite{halbertal2021moire,alden2013strain,halbertal2022unconventional}.

Similar strain-induced quasi-1D patterns have been also observed in, for example, TBG strained by a substrate \cite{mendoza2021strain}, large-angle twisted bilayer WTe$_2$ \cite{yang2025intrinsic}, and exfoliated
highly oriented pyrolytic graphite \cite{boi2025quasi}. 

\subsubsection{Giant atomic swirl}

\begin{figure}[t!]
\centering 
    \includegraphics[width=\linewidth]{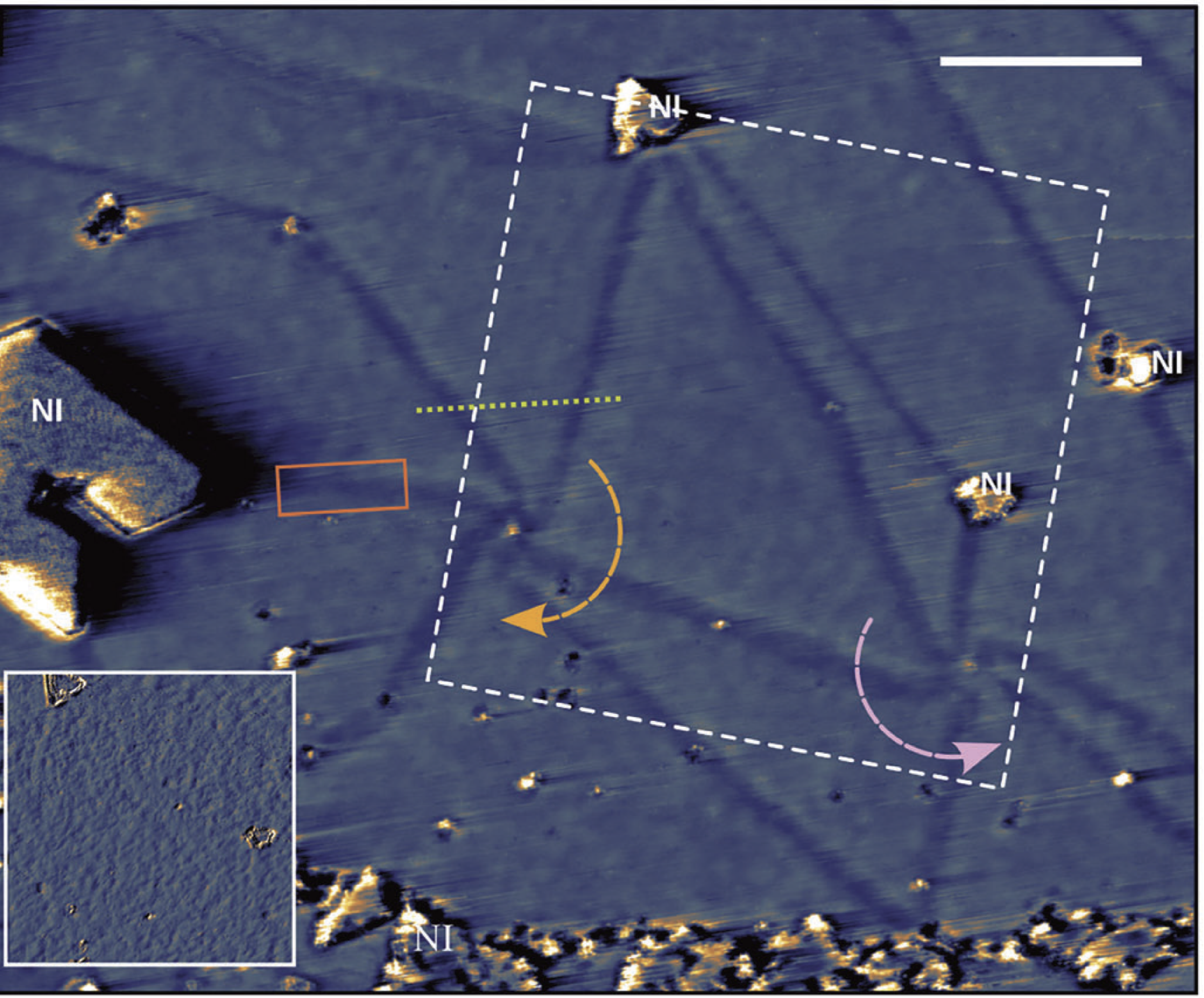}
    \caption{Strain induced giant swirl in bilayer graphene. STM current image ($V_b=-300$ mV, $I_t=250$ pA) for an intercalated surface of bilayer graphene on SiC, with a deformed moir\'e showing an (anti-) clockwise swriling feature in (blue) orange. NI indicate the non-intercalated regions. The inset is the STM image of the white dotted square region measued at $V_b=900$ mV, $I_t=250$ pA. The scal bar is 100 nm. Adapted under the terms of the CC BY license from Ref. \cite{mesple2023giant}. Copyright (2023) The author(s).}
\label{figs4_swirl}
\end{figure}

As noted in Section \ref{sec:lattice_relaxation}, relaxation-induced giant atomic swirls can emerge in the domain wall networks of moir\'e materials \cite{mesple2023giant,shi2025spontaneous}. Figure \ref{figs4_swirl} shows a moir\'e pattern formed in bilayer graphene with only biaxial heterostrain, which was produced by hydrogen intercalation of graphene on SiC \cite{mesple2023giant}. The strain was only imposed to the bottom graphene layer by the SiC reconstruction. Due to the presence of lattice reconstruction in the moir\'e pattern, domain walls connecting AA stacking formed a triangular network of AB/BA stackings. Notably, giant atomic swirls appeared spontaneously around the AA regions. The neighboring twirl chiralities tend to anti-align (blue and orange arrows). Similar flower-shaped domain walls were already reported in trilayer epitaxial graphene on silicon carbide \cite{lalmi2014flower}.

\section{Conclusions and outlook}

The addition of strain to twisted moir\'e materials greatly increases the possible moir\'e geometries and emergent electronic properties. Due to the magnifying effect of the moir\'e, significant changes in the system properties can be achieved even with marginal lattice deformations. On the one hand, this makes strain effects in moir\'e materials much more relevant than in isolated two-dimensional layers, which require significantly larger strain to change its properties. On the other hand, it means that a precise control of externally induced strain becomes crucial. As we emphasize in this review, the reward can be a powerful knob to tune the properties of moir\'e materials. 

While the geometrical properties of twisted and strained moir\'e patterns are by now relatively well understood, not much is yet known about the impact on the electronic properties. If anything, the strain adds to the complexity of understanding the rich correlation physics of twisted moir\'e materials. Useful approximations, as those resorting on symmetries, are likely to become invalid in the presence of strain, thus requiring more general schemes to tackle the problem. Although strain effects may seem unfavorable for correlations, they could actually facilitate or stabilize particular broken-symmetry phases. In either case, we anticipate that novel electronic properties in moir\'e heterostructures are likely to emerge as strain engineering techniques evolve.

\section{Acknowledgements}
We acknowledge support from NOVMOMAT, Grant PID2022-142162NB-I00 funded by MCIN/AEI/ 10.13039/501100011033 and, by “ERDF A way of making Europe”, and from the ‘Severo Ochoa’ Programme for Centres of Excellence in R\&D (CEX2020-001039-S/AEI/10.13039/501100011033). F.E. acknowledges support funding from the European Union's Horizon 2020 research and innovation programme under the Marie Skłodowska-Curie grant agreement No 101210351. F.G acknowledges the support from the  Department of Education of the Basque Government through the project No. \verb|PIBA\2023\1\0007(STRAINER)|. Z.Z. acknowledges support funding from the European Union's Horizon 2020 research and innovation programme under the Marie Skłodowska-Curie grant agreement No 101034431.

\bibliography{References_Revision}

\end{document}